\documentclass[twocolumn,aps,amsmath,amssymb,floatfix,superscriptaddress,prb,longbibliography]{revtex4-1}
\usepackage{graphicx}% Include figure files
\usepackage{dcolumn}% Align table columns on decimal point
\usepackage{bm}% bold math
\usepackage{xspace}
\usepackage{hyperref}
\usepackage{color}
\usepackage[flushleft]{threeparttable}

\def\tszo{Tm$_3$Sb$_3$Zn$_2$O$_{14}$\xspace}
\def\tsmo{Tm$_3$Sb$_3$Mg$_2$O$_{14}$\xspace}
\def\lszo{La$_3$Sb$_3$Zn$_2$O$_{14}$\xspace}
\def\lsmo{La$_3$Sb$_3$Mg$_2$O$_{14}$\xspace}
\begin{document}

%\setlength{\topmargin}{0in} %%%set margin to be zero to have proper ps printout
%%%when uploaded to arxiv and aps, this has to be commented.
\title{Disorder-induced spin-liquid-like behavior in kagome-lattice compounds}

\author{Zhen~Ma}
\affiliation{National Laboratory of Solid State Microstructures and Department of Physics, Nanjing University, Nanjing 210093, China}
\affiliation{Institute for Advanced Materials, Hubei Normal University, Huangshi 435002, China}
\author{Zhao-Yang~Dong}
\affiliation{Department of Applied Physics, Nanjing University of Science and Technology, Nanjing 210094, China}
\author{Si~Wu}
\author{Yinghao~Zhu}
\affiliation{Institute of Applied Physics and Materials Engineering, University of Macau, Avenida da Universidade, Taipa, 999078 Macau, China}
\author{Song~Bao}
\author{Zhengwei~Cai}
\affiliation{National Laboratory of Solid State Microstructures and Department of Physics, Nanjing University, Nanjing 210093, China}
\author{Wei~Wang}
\affiliation{National Laboratory of Solid State Microstructures and Department of Physics, Nanjing University, Nanjing 210093, China}
\affiliation{School of Science, Nanjing University of Posts and Telecommunications, Nanjing 210023, China}
\author{Yanyan~Shangguan}
\affiliation{National Laboratory of Solid State Microstructures and Department of Physics, Nanjing University, Nanjing 210093, China}
\author{Jinghui~Wang}
\author{Kejing~Ran}
\affiliation{National Laboratory of Solid State Microstructures and Department of Physics, Nanjing University, Nanjing 210093, China}
\affiliation{School of Physical Science and Technology and ShanghaiTech Laboratory for Topological Physics, ShanghaiTech University, Shanghai 200031, China}
\author{Dehong~Yu}
\author{Guochu~Deng}
\author{Richard~A.~Mole}
\affiliation{Australian Nuclear Science and Technology Organisation (ANSTO), New Illawarra Road, Lucas Heights, New South Wale 2234, Australia}
\author{Hai-Feng~Li}
\email{haifengli@umac.edu.mo}
\affiliation{Joint Key Laboratory of the Ministry of Education, Institute of Applied Physics and Materials Engineering, University of Macau, Avenida da Universidade, Taipa, Macau SAR 999078, China}
\author{Shun-Li Yu}
\email{slyu@nju.edu.cn}
\author{Jian-Xin Li}
\email{jxli@nju.edu.cn}
\author{Jinsheng Wen}
\email{jwen@nju.edu.cn}
\affiliation{National Laboratory of Solid State Microstructures and Department of Physics, Nanjing University, Nanjing 210093, China}
\affiliation{Collaborative Innovation Center of Advanced Microstructures, Nanjing University, Nanjing 210093, China}

%\date{\today}

\begin{abstract}
Quantum spin liquids (QSLs) are an exotic state of matter that is subject to extensive research. However, the relationship between the ubiquitous disorder and the QSL behaviors is still unclear. Here, by performing comparative experimental studies on two kagom\'{e}-lattice QSL candidates, \tszo and \tsmo, which are isostructural to each other but with strong and weak structural disorder, respectively, we show unambiguously that the disorder can induce spin-liquid-like features. In particular, both compounds show dominant antiferromagnetic interactions with a Curie-Weiss temperature of -17.4 and -28.7~K for \tszo and \tsmo, respectively, but remain disordered down to about 0.05~K. Specific heat results suggest the presence of gapless magnetic excitations characterized by a residual linear term. Magnetic excitation spectra obtained by inelastic neutron scattering (INS) at low temperatures display broad continua. All these observations are consistent with those of a QSL. However, we find in \tszo which has strong disorder resulting from the random mixing of the magnetic Tm$^{3+}$ and nonmagnetic Zn$^{2+}$, that the low-energy magnetic excitations observed in the specific heat and INS measurements are substantially enhanced, compared to those of \tsmo which has much less disorder. We believe that the effective spins of the Tm$^{3+}$ ions in the Zn$^{2+}$/Mg$^{2+}$ sites give rise to the low-energy magnetic excitations, and the amount of the random occupancy determines the excitation strength. These results provide direct evidence of the mimicry of a QSL caused by disorder.
\end{abstract}

\maketitle

\section{Introduction}

Quantum spin liquids (QSLs) represent a novel state of matter in which spins are highly entangled, but do not order nor freeze even in the zero-temperature limit~\cite{Anderson1973153,nature464_199}. Such a state does not involve any spontaneous symmetry breaking, which is beyond Landau's paradigm for a phase and the associated transition~\cite{imai2016quantum}. They are proposed to host fractional excitations and emergent gauge structures, and thus are promising candidates for quantum computation~\cite{Kitaev20032,Barkeshli722}. Furthermore, high-temperature superconductivity may emerge from carrier doping a QSL~\cite{anderson1,Baskaran1987973,lee:17}. Thus, it has been a long-sought goal to achieve the QSL state. However, spins often tend to order at low temperatures~\cite{Neel1985}. One approach to the goal is to introduce geometrical frustration into a low-spin system to enhance quantum fluctuations, so magnetic exchange interactions cannot be satisfied simultaneously among different lattice sites and the static magnetic order is prohibited~\cite{Anderson1973153,arms24_453}. By now, a number of QSL candidates resulting from geometrical frustrations have been proposed and explored experimentally~\cite{0034-4885-80-1-016502,RevModPhys.89.025003,Wen2019Experimental}, and some typical examples include organic triangular-lattice systems $\kappa$-(ET)$_2$Cu$_2$(CN)$_3$~\cite{PhysRevLett.91.107001,PhysRevLett.95.177001,Ohira2006,np4_459,np5_44,Furukawa2018Quasi} and EtMe$_3$Sb[Pd(dmit)$_2$]$_2$~\cite{np6_673,Yamashita1246,nc2_275}, kagom\'e-lattice compound ZnCu$_3$(OH)$_6$Cl$_2$~\cite{prl98_107204,nature492_406,RevModPhys.88.041002}, inorganic triangular-lattice compound YbMgGaO$_4$~\cite{sr5_16419,prl115_167203,np13_117,nature540_559,PhysRevLett.117.097201} and its sister compound YbZnGaO$_4$~\cite{PhysRevLett.120.087201}, and more recently found triangular-lattice system delafossites~\cite{Liu_2018,PhysRevB.98.220409,np15_1058,PhysRevB.100.144432,PhysRevB.100.241116,PhysRevB.100.224417,doi:10.1021/acsmaterialslett.9b00464,PhysRevMaterials.3.114413,PhysRevB.100.220407}. The disorder-free delafossites with effective spin-1/2 moments provide an excellent platform to unveil the QSL nature in a clean system. However, for most of these QSL candidates, the magnetic or nonmagnetic disorder can be significant, and complicates the interpretation of the intrinsic physics of the investigated systems~\cite{PhysRevB.94.060409,PhysRevLett.119.157201,PhysRevB.97.184413,np13_117,PhysRevLett.118.107202,PhysRevLett.118.087203,PhysRevLett.120.087201,PhysRevLett.120.207203,nc_Itamar,PhysRevX.8.031001,PhysRevX.8.031028,PhysRevX.8.041040,PhysRevLett.123.087201,PhysRevB.96.174418,PhysRevLett.115.077001}. For this reason, how disorder affects the QSL state is still a controversial issue.

Recently, $RE_3$Sb$_3M_2$O$_{14}$ as a new family of two-dimensional kagom\'e-lattice compounds were synthesized~\cite{doi:10.1002/pssb.201600256,Sanders2016RE,PhysRevB.95.104439}, where $RE^{3+}$ represents rare-earth ions and $M^{2+}$ denotes nonmagnetic Zn$^{2+}$ or Mg$^{2+}$ ions. Among these compounds, Tm$_3$Sb$_3$Zn$_2$O$_{14}$ was proposed to be a QSL~\cite{PhysRevB.98.174404}. The Tm$^{3+}$ ion has an electron configuration of $4f^{12}$ and, according to the Hund's rule, it has a total angular momentum of $J = 6$ with a 13-fold degeneracy. Considering the crystal-electric-field~(CEF) effect, the degeneracy of $J$ will be lifted, which was suggested to give rise to a non-Kramers doublet ground state~\cite{PhysRevB.98.174404}. In \ref{subseccef}, we will discuss the ground state and CEF excitations in more details. There is no signature of magnetic phase transition by magnetic susceptibility, heat capacity, and muon-spin relaxation measurements, suggesting the possible realization of a gapless QSL ground state~\cite{PhysRevB.98.174404}. However, it was shown that there was a significant site mixing between the magnetic Tm$^{3+}$ and nonmagnetic Zn$^{2+}$ sites~\cite{PhysRevB.98.174404}, causing a strong disorder effect that could impact the proposed QSL state.

\begin{figure*}[htb]
\centerline{\includegraphics[width=0.8\linewidth]{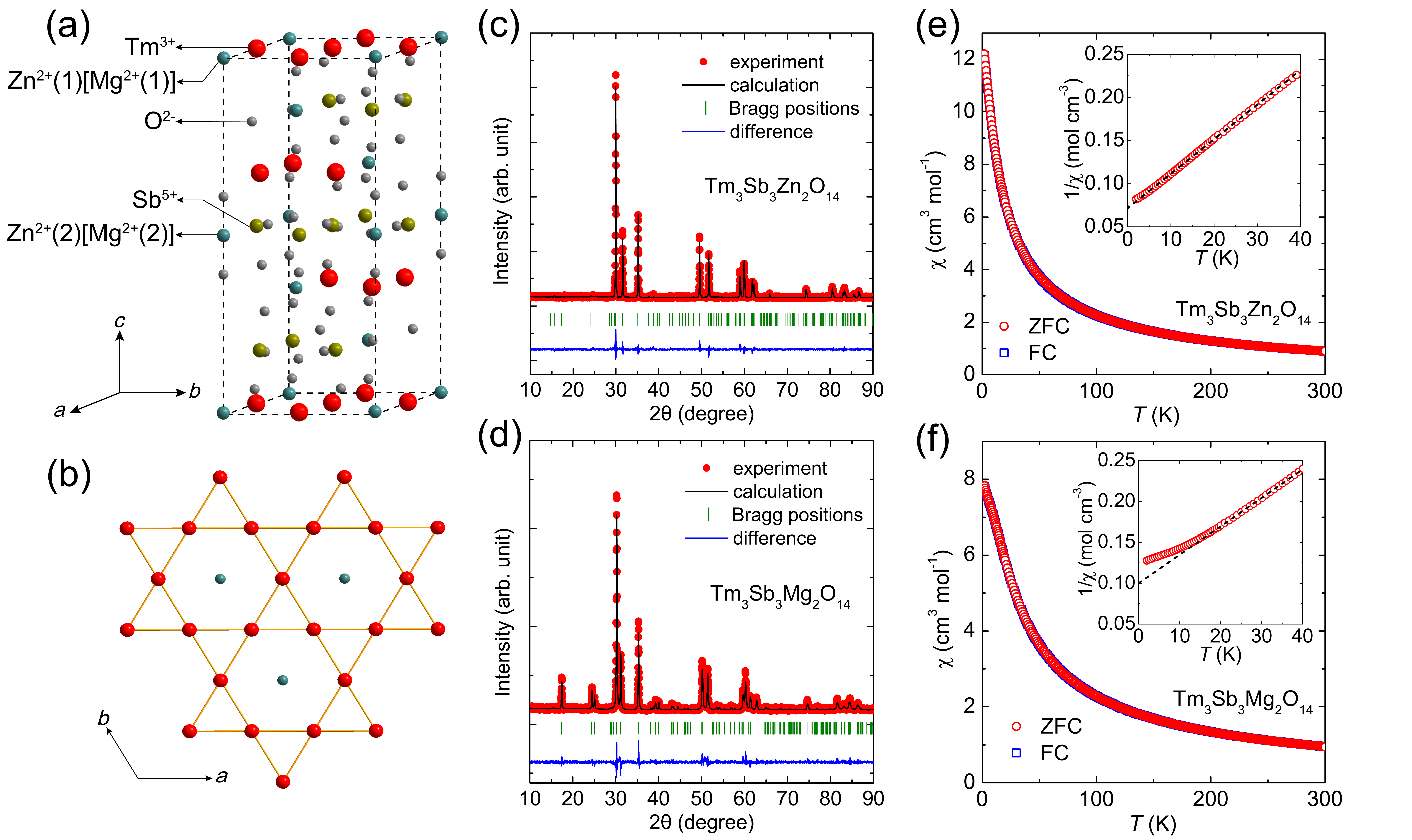}}
\caption{
(a) Schematic crystal structure of \tszo and \tsmo with $R\bar3m$ space group. (b) Top view of the kagom\'e layer consisting of Tm$^{3+}$ ions at the corners and Zn$^{2+}$ ions at the centers. (c) and (d) Rietveld refinement results for the powder X-ray diffraction data for \tszo and \tsmo, respectively. (e) and (f) Temperature dependence of magnetic susceptibility for \tszo and \tsmo, respectively, measured under zero-field-cooling (ZFC) and field-cooling (FC) conditions with a magnetic field of 0.1~T from 2 to 300~K. The insets of (e) and (f) show the inverse susceptibility data at low temperatures. The dashed lines are the fits with the Curie-Weiss law.
\label{fig1}}
\end{figure*}

\begin{table*}[htb]
  \begin{threeparttable}
\caption{Room-temperature XRD pattern refinements for \tszo and \tsmo.}
\label{tab:para}
\begin{tabular*}{\textwidth}{@{\extracolsep{\fill}}ccccccccccc}
\hline \hline
\begin{minipage}{2cm}\vspace{1mm} Compound \vspace{1mm} \end{minipage} & Atom & Wyckoff position & $x$ & $y$ & $z$ & Occ. & $U_{\rm iso}$~(\AA$^2$) & $a$~(\AA) & $c$~(\AA) & $\chi^2$\\
\hline
\begin{minipage}{2cm}\vspace{1mm}  \vspace{1mm} \end{minipage} & Tm & 9e & 0.5 & 0 & 0 & 0.65(3) & & & & \\
  & Zn(disorder) & 9e & 0.5 & 0 & 0 & 0.35(3) & & & & \\
  & Zn(1) & 3a & 0 & 0 & 0 & 0.45(3) & & & & \\
  & Tm(disorder) & 3a & 0 & 0 & 0 & 0.55(3) & & & & \\
 \tszo & Zn(2) & 3b & 0 & 0 & 0.5 & 0.44(4) & 0.025 & 7.3541(2) & 16.9956(5) & 3.84 \\
 & Tm(disorder) & 3b & 0 & 0 & 0.5 & 0.56(4) & & & & \\
  & Sb & 9d & 0.5 & 0 & 0.5 & 1 & & & & \\
  & O(1) & 6c & 0 & 0 & 0.393(6) & 1 & & &  & \\
  & O(2) & 18h & 0.504(3) & -0.504(3) & 0.117(2) & 1 & & & & \\
  & O(3) & 18h & 0.143(3) & -0.143(3) & -0.026(2) & 1 & & & & \\
\hline
\begin{minipage}{2cm}\vspace{1mm}  \vspace{1mm} \end{minipage} & Tm & 9e & 0.5 & 0 & 0 & 0.93(1) & & & & \\
  & Mg (disorder) & 9e & 0.5 & 0 & 0 & 0.07(1) & & &  & \\
  & Mg(1) & 3a & 0 & 0 & 0 & 0.82(1) & & & & \\
  & Tm(disorder) & 3a & 0 & 0 & 0.5 & 0.18(1) & & &  & \\
  \tsmo & Mg(2) & 3b & 0 & 0 & 0.5 & 0.95(1) & 0.025 & 7.2772(5) & 17.2278(1) & 3.67 \\
  & Tm(disorder) & 3b & 0 & 0 & 0.5 & 0.05(1) & & &  & \\
  & Sb & 9d & 0.5 & 0 & 0.5 & 1 & & & &  \\
  & O(1) & 6c & 0 & 0 & 0.615(2) & 1 & & &  & \\
  & O(2) & 18h & 0.522(2) & -0.522(2) & 0.139(1) & 1 & & &  & \\
  & O(3) & 18h & 0.132(2) & -0.132(2) & -0.058(1) & 1 & & &  & \\
\hline \hline
\end{tabular*}
\begin{tablenotes}
      \small
      \item $T$ = 300~K, wavelength of the x-ray $\lambda$ = 1.54~\AA, and space group: $R\bar3m$.\\
      \item $U_{\rm iso}$ denotes the isotropic actomic displacement from the equilibrium positions, and $\chi^2$ represents the goodness of fitting.\\
\end{tablenotes}
\end{threeparttable}
\end{table*}

In this work, in order to identify the role of disorder in the QSL candidates, we choose \tszo and its sister compound \tsmo with quantifiable disorder, that is, strong and weak disorder in the former and latter, respectively, and investigate how disorder affects the material's magnetic properties. On one hand, the absence of magnetic order and spin freezing down to $\sim$50~mK far below the Curie-Weiss temperature ($\Theta_{\rm CW}$), the presence of residual linear term in the specific heat, and the observation of broad gapless magnetic excitations resemble those of gapless QSLs. On the other hand, we find that the disorder resulting from the random mixing of magnetic Tm$^{3+}$ and nonmagnetic Zn$^{2+}$ or Mg$^{2+}$ in the Tm layers is intimately correlated with the strength of the low-energy magnetic excitations. In particular, in \tszo which has strong disorder, as identified from the structural refinement and CEF excitations, the value of the residual linear term in the specific heat is about 4 times larger, and the intensity of the gapless spin excitations in the INS spectra is greatly enhanced, compared to those of \tsmo with much less disorder. These results demonstrate conclusively that disorder in a geometrically frustrated compound can make it mimic a QSL.

\section{Experimental Details}

Polycrystalline samples of \tszo (\lszo) and \tsmo (\lsmo) were synthesized by conventional solid-state reactions with stoichiometric amounts of Tm$_2$O$_3$ (99.99\%) (Lu$_2$O$_3$, 99.99\%), Sb$_2$O$_3$ (99.99\%), ZnO (99.99\%), and MgO (99.99\%) powders. The mixtures of the precursor compounds of \tszo (\lszo) were carefully ground and sintered at 1200~$^{\circ}$C for 3 days. For \tsmo (\lsmo) samples, a higher reaction temperature of 1350~$^{\circ}$C and longer reaction time of 5 days were required to obtain the pure phase. X-ray diffraction (XRD) data were collected in an x-ray diffractometer (X$^\prime$TRA, ARL) using the Cu-$K_\alpha$ edge with a wavelength of 1.54~\AA. Rietveld refinements on the data were run by the EXPGUI and GSAS programs~\cite{Toby:hw0089,Toby:aj5212}. The dc magnetic susceptibility was measured in a Quantum Design physical property measurement system (PPMS, EverCool). Specific heat above 2~K was measured on 5.6-mg \tszo and 4.7-mg \tsmo samples, respectively, in a PPMS EverCool. The data below 2~K were collected on 1.7-mg \tszo and 8.0-mg \tsmo in a PPMS DynaCool equipped with a dilution refrigerator.

INS experiments on 5.3-g \tszo and 4.5-g \tsmo polycrystalline samples were carried out on PELICAN equipped with a dilution refrigerator, a cold neutron time-of-flight (TOF) spectrometer located at ANSTO at Lucas Heights, Australia. The powders were loaded into a pure copper can in the dilution refrigerator, which was able to cool down to around 50~mK. The incident neutron wavelength was selected as $\lambda~\sim$ 4.69 \AA, corresponding to an incident energy of 3.69~meV and an energy resolution of $\Delta E$ = 0.067~meV (half width at half maximum, HWHM). For each temperature, we collected data for about 12 hours. The CEF experiments were performed on 8-g powders on a thermal triple-axis spectrometer TAIPAN at ANSTO. The powders were loaded into an aluminum can and then mounted onto a closed-cycle refrigerator which could reach 1.6~K. A pyrolytic graphite (PG) filter was placed after the sample to reduce contaminations from higher-order neutrons. The beam collimations were 0$^\prime$-40$^\prime$-sample-40$^\prime$-0$^\prime$. A fixed-final-energy ($E_{\rm{f}}$) mode with $E_{\rm{f}}$ = 14.87~meV was used in the measurements. The resulting energy resolution was about 0.41~meV (HWHM). Measurements were performed under vertical-focusing conditions for both the monochromator and analyzer on TAIPAN.

\section{Results}

\subsection{Structure and magnetic susceptibility}

Figures~\ref{fig1}(a) and~\ref{fig1}(b) show the schematics of the crystal structure and the two-dimensional kagom\'e layer of \tszo, respectively. Magnetic Tm$^{3+}$ ions forming corner-shared kagom\'e-lattice layers are well separated by nonmagnetic layers and have an ABC stacking arrangement along the $c$ axis~\cite{PhysRevB.95.104439,PhysRevB.98.174404}. We have performed Rietveld refinements on the XRD data of \tszo and the results are shown in Fig.~\ref{fig1}(c). For the $R\bar3m$ space group with a perfect kagom\'{e} lattice, there should be some Bragg reflections below 30 degrees, which are absent in our data. In order to capture the absence of low-angle reflections, we have to allow some mixings between the Tm$^{3+}$ and Zn$^{2+}$(1) sites. Such a site-mixing model was used in Ref.~\onlinecite{PhysRevB.98.174404} to analyze the XRD data in \tszo as well. In order to keep the stoichiometry, we need to free the site mixing between the Tm$^{3+}$ and Zn$^{2+}$(2) sites as well, which was not considered in Ref.~\onlinecite{PhysRevB.98.174404}. The detailed refinement parameters are listed in Table~\ref{tab:para}. Our refinement results show that there are around 35\% Zn$^{2+}$ ions occupying Tm$^{3+}$ positions, and 55\% and 56\% Tm$^{3+}$ ions occupying Zn$^{2+}$(1) and Zn$^{2+}$(2) positions, respectively. The strong site mixing of Tm$^{3+}$ and Zn$^{2+}$ reduces the distinctness of their original positions and then increases the crystallographic symmetry, {\it i.e.}, from kagom\'{e} to triangular within the $a$-$b$ plane~\cite{PhysRevB.95.104439,PhysRevB.98.174404}. This naturally explains the absence of the low-angle Bragg reflections in Fig.~\ref{fig1}(c).

The random site mixing between the magnetic and nonmagnetic sites is expected to have a strong impact on the magnetic properties, which is indeed the case as will be discussed later. To address this issue, it is better to have a comparative compound that is isostructural to \tszo but has less disorder resulting from the random mixing of the magnetic and nonmagnetic ions. For this purpose, we have replaced the nonmagnetic Zn in \tszo with Mg and synthesized \tsmo, which has less disorder as we show below.  We have performed similar refinements and the results are shown in Fig.~\ref{fig1}(d) and Table~\ref{tab:para}. The XRD pattern for \tsmo shown in Fig.~\ref{fig1}(d) is almost the same as that for \tszo shown in Fig.~\ref{fig1}(c), except for the additional reflections below 30 degree, which is expected for the $R\bar3m$ space group with perfect kagom\'{e} layers. As shown in Table~\ref{tab:para}, the crystal structure for both compounds is the same, but the amount of Tm$^{3+}$ in the Mg$^{2+}$ sites is significantly reduced. We believe that the different amounts of disorder in \tszo and \tsmo are due to the different radii of the Zn$^{2+}$ and Mg$^{2+}$ ions. Compared to the smaller Mg$^{2+}$, the larger Zn$^{2+}$ ions are closer to Tm$^{3+}$ ions in size, and thus it is easier to occupy each other randomly. In support of this point, it has been reported that when the radius of the rare-earth ion becomes large enough to have an obvious difference from that of Zn$^{2+}$ ion, such as Dy$^{3+}$ or larger ones in the 4$f$ row, the disorder effect will be weakened significantly~\cite{PhysRevB.95.104439}.

We further characterize both compounds by measuring the magnetic susceptibility ($\chi$)  with a magnetic field of 0.1~T, and the results are shown in Figs.~\ref{fig1}(e) and~\ref{fig1}(f). There is no indication of long-range magnetic order down to 2~K. In addition, the absence of the bifurcation of zero-field-cooling (ZFC) and field-cooling (FC) data also indicates that there is no spin freezing at the lowest temperature measured. The inverse susceptibility is linear for most of the temperature range, except for the slight deviation at low temperatures, as shown in the inset of Figs.~\ref{fig1}(e) and~\ref{fig1}(f). Such a deviation may be associated with the development of short-range magnetic correlations, and is commonly observed in QSL candidates~\cite{PhysRevB.76.132411,PhysRevLett.120.087201,prl115_167203,PhysRevB.95.174414}. It can also be explained with the thermal population of CEF levels and we will discuss it later. From the Curie-Weiss fits, we obtain the Curie-Weiss temperature $\Theta_{\rm {CW}}$ of -17.4 and -28.7~K for \tszo and \tsmo, respectively, implying dominating antiferromagnetic spin correlations in these materials. These results are consistent with the QSL state expected for a kagom\'{e}-lattice system with antiferromagnetic interactions and strong geometrical frustration~\cite{nature464_199,Wen2019Experimental,PhysRevB.98.174404}. On the other hand, the disorder resulting from the random mixing of the magnetic and nonmagnetic sites discussed earlier can be also responsible for these observations~\cite{PhysRevLett.120.087201,PhysRevLett.119.157201,PhysRevX.8.031028,PhysRevX.8.041040}.

\subsection{CEF excitations}\label{subseccef}

\begin{figure}[htb]
\centerline{\includegraphics[width=3.45in]{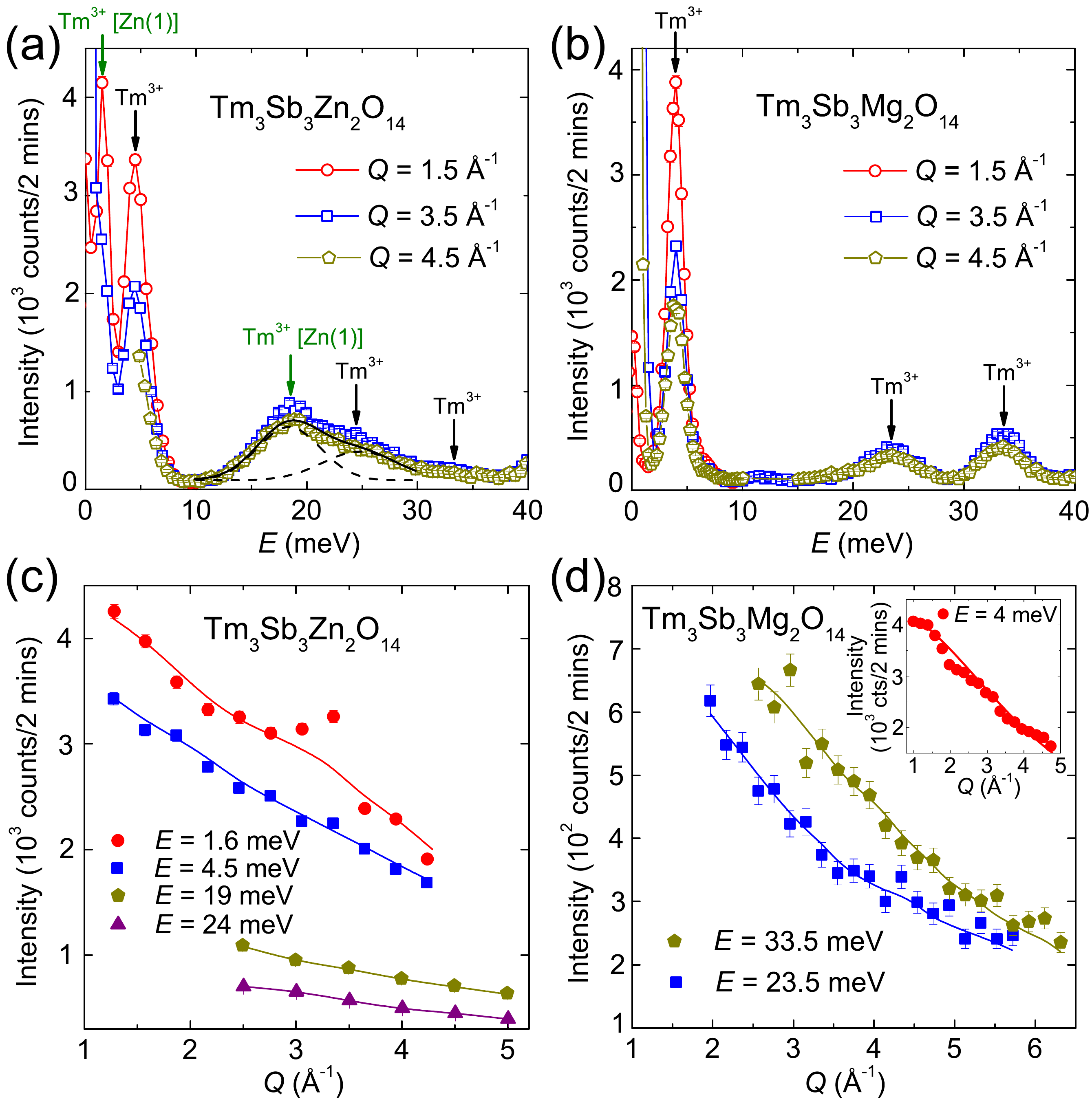}}
\caption{(a) and (b) Crystal-electric-field (CEF) excitations below 40~meV for \tszo and \tsmo, respectively, measured at 1.6~K on TAIPAN spectrometer. Green and black arrows in (a) indicate the positions of CEF transitions at 1.6 and 18.5~meV caused by Tm$^{3+}$ at the Zn(1) sites, and 4.5, 24.8, and 33.5~meV caused by Tm$^{3+}$ at the original sites for \tszo. Black solid line represents the convolution with two Gaussian functions denoted by dashed lines. Arrows in (b) indicate the positions of CEF transitions at 4, 23.5, and 33.5~meV for \tsmo. (c) and (d) $\bm{Q}$-dependence intensities of the CEF levels for \tszo and \tsmo, respectively. Solid lines indicate that the intensities of the CEF excitations follow the magnetic form factor of Tm$^{3+}$ ion well with $I~\propto~|f(\bm{Q})|^2$. $I$ and $|f(\bm{Q})|$ denote excitation intensity measured and magnetic form factor of Tm$^{3+}$ ion, respectively. Errors represent one standard deviation throughout the paper.
\label{fig2}}
\end{figure}

To further investigate the disorder effect in these two compounds, we measured the CEF excitations, and the experimental results are shown in Fig.~\ref{fig2}. The experimental CEF spectra of \tsmo are shown in Fig.~\ref{fig2}(b). There are three well-isolated CEF transitions centered at 4, 23.5, and 33.5~meV, respectively, with no asymmetry nor significant broadening observed. For \tszo, there are two CEF levels observed at 1.6 and 4.5~meV. Both peaks are slightly broader than the instrument resolution of 0.41~meV but still relatively sharp. However, another one centered at 19~meV is asymmetric and much broader. To analyze the experimental results in \tszo, we need to consider the random site mixing between the Tm$^{3+}$ and Zn$^{2+}$ ions, which is strong as shown in Table~\ref{tab:para}. A Zn$^{2+}$(1) cation with surrounding six O$^{2-}$ anions produces a ZnO$_6$ polyhedra, while the Zn$^{2+}$(2) and Tm$^{3+}$ cations are both surrounded by eight O$^{2-}$ anions~\cite{doi:10.1002/pssb.201600256}. As a result, the Tm$^{3+}$ ions at the original and Zn(1) sites have different ligand environments and should give rise to two different sets of CEF excitations associated with these two sites. This indeed makes the CEF pattern of \tszo more complicated than that of \tsmo. By comparing their experimental results in Figs.~\ref{fig2}(a) and~\ref{fig2}(b), we believe the CEF transitions observed at 4.5, 24.8, and 33.5~meV of \tszo can be attributed to excitations of the Tm$^{3+}$ ions at the original sites, since these energies are almost the same as those of \tsmo, although the one at 33.5~meV is weaker than that in the latter. In \tszo, the remaining two CEF levels at 1.6 and 18.5~meV which are completely absent in \tsmo should result from the CEF excitations of Tm$^{3+}$ cations at the Zn(1) sites. In this case, the asymmetric and very broad peak around 19~meV for \tszo shown in Fig.~\ref{fig2}(a) is actually composed of two CEF levels centered at 18.5 and 24.8~meV resulting from two different sites. These results show clearly that the site mixing between the magnetic and nonmagnetic ions will have a strong impact on the CEF excitations.

In order to confirm the signals we have observed here indeed originate from CEF excitations, some $\bm{Q}$ scans located at representative energy levels were performed. In Figs.~\ref{fig2}(c) and~\ref{fig2}(d), the behavior of monotonic decrease of intensities with increasing $\bm{Q}$ follows the magnetic form factor of a Tm$^{3+}$ ion, distinct from phonon scatterings. Moreover, INS results of a nonmagnetic reference compound La$_3$Sb$_3$Mg$_2$O$_{14}$ show phonon scattering is only pronounced at a larger range of $\bm{Q}$ $\ge$ 10~\AA$^{-1}$~(Ref.~\onlinecite{PhysRevB.98.134401}), which is far away from our currently investigated area.

The rare-earth ions in the materials are strongly influenced by the electrostatic environment they occupy. Therefore, in order to quantitatively identify how the $(2J+1)$-fold ($J=6$) spin-orbital degeneracy is lifted by the CEF effect, a CEF analysis was performed, and the effective Hamiltonian was obtained by the point-charge model according to the point-group symmetry\cite{HUTCHINGS1964227},
\begin{equation}\label{CEF}
  H_{\rm CEF}=\sum_{n,m}B_n^m O_n^m,
\end{equation}
where $O_n^m$ and $B_n^m$ are the Steven operators\cite{Stevens_1952} and CEF parameters, respectively. In the following, $B_n^m $ are calculated based on the 15 nontrivial CEF parameters shown in Table~\ref{CEFp} within the point-charge approximation\cite{HUTCHINGS1964227,PhysRevB.98.134401}.

Since the CEF of Tm$^{3+}$ at the original sites has a very low symmetry of the $C_{2h}$ point group with eight oxygen ligands, the $13$-fold degeneracy of the ground-state manifold $^3H_6$ is expected to be completely lifted. Fortunately, the diagonalization of Eq.~(\ref{CEF}) predicts that the splitting of two low-lying singlets are small enough to be regarded as a nearly degenerate non-Kramers doublet. The effective spin in the doublet is an easy-axis moment, and the components are mostly $J_z=\pm6$ along the easy axis.
If we choose our local axes such that the two-fold rotation symmetry is about the $y$ axis and the easy axis is the $z$ axis, the doublet shown in Table~\ref{doublet} can be well described by symmetric and antisymmetric combinations of $J_z=\pm6$ states:
\begin{equation}\label{state}
  |+\rangle\approx\frac{\sqrt{2}}{2}(|6\rangle+|-6\rangle),|-\rangle\approx\frac{\sqrt{2}}{2}(|6\rangle-|-6\rangle).
\end{equation}

Due to the low symmetry, the CEF peaks observed experimentally are not sufficient to simulate all the CEF parameters. In addition, the disorder would further influence the simulation. In this case, we simply do the symmetry analysis and obtain the effective CEF Hamiltonian by point-charge approximation. Since the lifting of the degeneracy is mainly determined by the point-group symmetry, we think the analysis is sufficient to identify the low-lying states, which are very far away from other higher energy levels. Further fitting would revise the higher energy levels, but the components of the non-Kramers doublet would not change significantly. A similar non-Kramers doublet was also reported for Ho$_3$Sb$_3$Mg$_1$O$_{14}$~(Ref.~\onlinecite{PhysRevX.10.031069}) and Tb$_3$Sb$_3$Mg$_2$O$_{14}$~(Ref.~\onlinecite{dun2020effective}), which share the same structure. The fitting analysis of the same structure of other crystals also supported that the symmetry analysis is qualitatively efficient for the low-lying states\cite{PhysRevB.98.134401}.

For the Tm$^{3+}$ cations at the Zn($1$) sites in Tm$_3$Zn$_2$Sb$_3$O$_{14}$, the ligand environment is a squashed oxygen octahedron, whose symmetry is higher than the original sites. The CEF with $D_{3d}$ point group symmetry will split the $13$-degenerate states into five singlets and four doublets. It has been revealed that two low-lying singlets would compose the nearly degenerate non-Kramers doublet\cite{PhysRevX.10.011007}. The components of the nearly degenerate non-Kramers doublet ground state are primarily $J_z=\pm6$ and $\pm3$ states. Therefore, an effective spin-$1/2$ can be defined in this doublet\cite{PhysRevX.10.011007}. Due to the occupations of the Tm$^{3+}$ cations at the original and Zn(1) sites, the two different ligand environments lead to different CEF excitations, which results in a very different CEF pattern shown in Fig.~\ref{fig2}(a).

%\begin{table*}[htb]
  %\begin{threeparttable}
%\caption{The CEF parameters obtained from the point-charge approximation.}
%\label{CEFp}
%\begin{tabular*}{\textwidth}{@{\extracolsep{\fill}}cccccccc}
%\hline \hline
%\begin{minipage}{2cm}\vspace{1mm} $B_n^m$ \vspace{1mm} \end{minipage} & $B_2^0$ & $B_2^1$ & $B_2^2$ & $B_4^0$ & $B_4^1$ & $B_4^2$ & $B_4^3$  \\
%\hline
% meV & -1.40866 & 2.02733 & 0.79967 & -0.00252 & -0.01587 & 0.00610 & 0.03075  \\
%\hline \hline
%\begin{minipage}{2cm}\vspace{1mm} $B_4^4$ \vspace{1mm} \end{minipage} & $B_6^0$ & $B_6^1$ & $B_6^2$ & $B_6^3$ & $B_6^4$ & $B_6^5$ & $B_6^6$  \\
%\hline
% 0.02983 & -0.00002 & -0.00007 & 0.00004 & 0.00016 & -0.00017 & -0.00054 & -0.00005 \\
%\hline \hline
%\end{tabular*}
%\end{threeparttable}
%\end{table*}

\begin{table}[htb]
\caption{The CEF parameters obtained from the point-charge approximation.}
\label{CEFp}
\begin{tabular*}{\columnwidth}{@{\extracolsep{\fill}}cc}
\hline\hline
\begin{minipage}{2cm}\vspace{1mm} $B_n^m$ \vspace{1mm} \end{minipage}  &  \begin{minipage}{2cm}\vspace{1mm} meV \vspace{1mm} \end{minipage}  \\
\hline
\begin{minipage}{2cm}\vspace{0.5mm} $B_2^0$ \vspace{0.5mm} \end{minipage}  & -1.40866 \\
$B_2^1$   & 2.02733 \\
$B_2^2$   & 0.79967 \\
$B_4^0$ & -0.00252 \\
$B_4^1$ & -0.01587 \\
$B_4^2$ & 0.00610 \\
$B_4^3$ & 0.03075 \\
$B_4^4$ & 0.02983 \\
$B_6^0$ & -0.00002 \\
$B_6^1$ & -0.00007 \\
$B_6^2$ & 0.00004 \\
$B_6^3$ & 0.00016 \\
$B_6^4$ & -0.00017 \\
$B_6^5$ & -0.00054 \\
$B_6^6$ & -0.00005 \\
\hline\hline
\end{tabular*}
\end{table}

\begin{table*}[htb]
  \begin{threeparttable}
\caption{Eigenvalues and eigenvectors for the non-Kramers doublet ground state of Tm$^{3+}$ at the original sites. The first column indicates the energies and the rest indicate the antisymmetric and symmetric states of the doublet.}
\label{doublet}
\begin{tabular*}{\textwidth}{@{\extracolsep{\fill}}ccccccccccccccc}
\hline \hline
\begin{minipage}{2cm}\vspace{1mm} $E$ (meV) \vspace{1mm} \end{minipage} &  & $|-6\rangle$ & $|-5\rangle$ & $|-4\rangle$ & $|-3\rangle$ & $|-2\rangle$ & $|-1\rangle$ & $|0\rangle$ & $|1\rangle$ & $|2\rangle$ & $|3\rangle$ & $|4\rangle$ & $|5\rangle$ & $|6\rangle$ \\
\hline
 0.000 & $|-\rangle=$ & (0.701 & 0.010 & -0.079 & 0.042 & 0.008 & -0.024 & 0.000 & -0.024 & -0.008 & 0.042 & 0.079 & 0.010 & -0.701)\\
  0.018 & $|+\rangle=$ & (0.701 & 0.008 & -0.086 & 0.031 & 0.001 & -0.017 & 0.024 & 0.017 & 0.001 & -0.031 & -0.086 & -0.008 & 0.701)\\
\hline \hline
\end{tabular*}
\end{threeparttable}
\end{table*}

\subsection{Specific heat results}

\begin{figure}[htb]
\centerline{\includegraphics[width=3.45in]{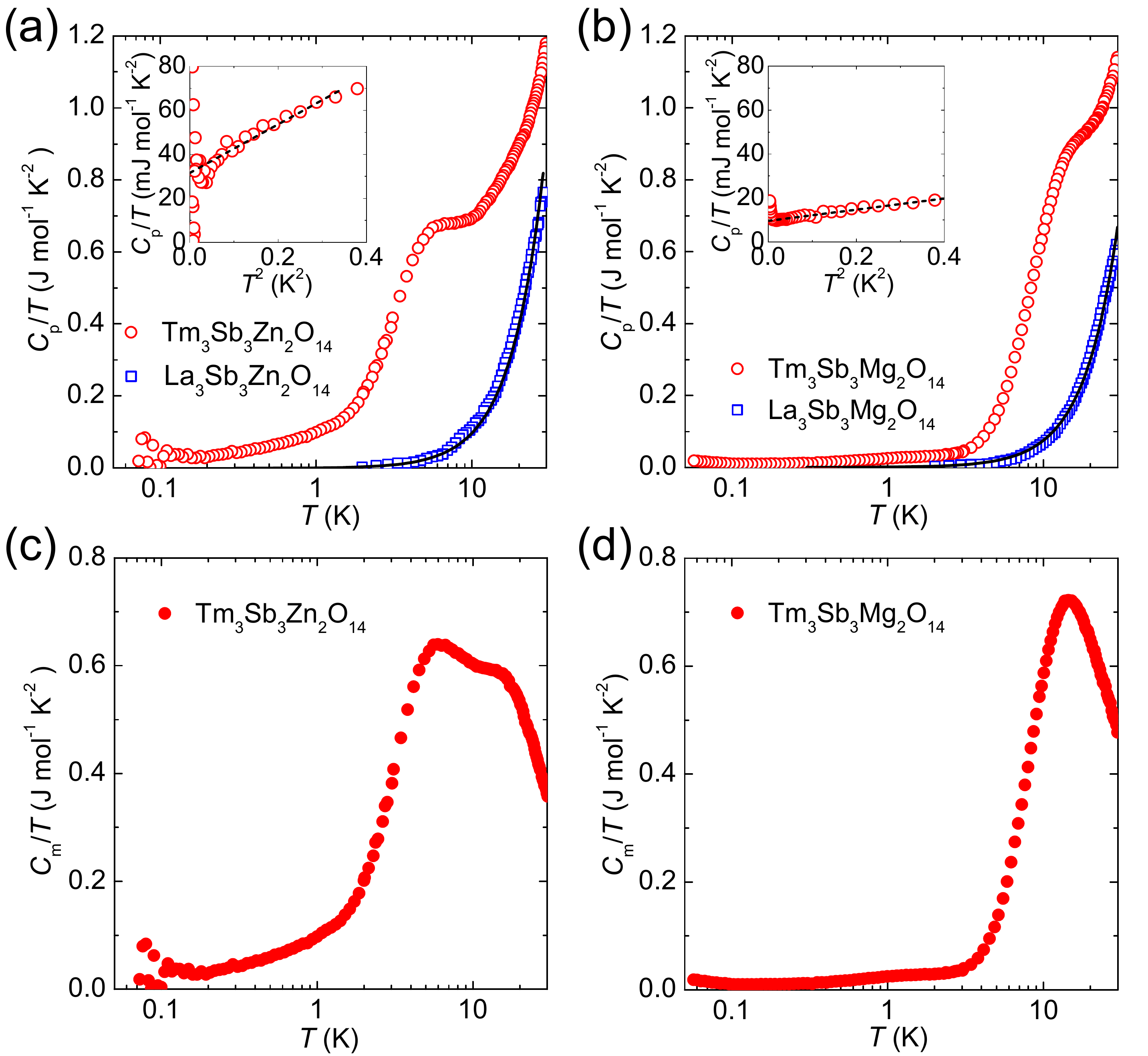}}
\caption{(a) and (b) Specific heat results of \tszo and \tsmo at ultralow temperatures. The specific heat of nonmagnetic references \lszo and \lsmo are also shown. Solid lines are the fits with Debye model as $C_{\rm p}~\sim~T^3$ for the nonmagnetic compounds. Insets show the low-temperature $C_{\rm p}/T$ $vs.$ $T^2$. Dashed lines are the linear fits. (c) and (d) Magnetic specific heat~($C_{\rm m}$) results of \tszo and \tsmo after subtracting the contribution from the lattice using nonmagnetic reference compounds \lszo and \lsmo, respectively.
\label{fig3}}
\end{figure}

We performed ultralow-temperature specific heat ($C_{\rm p}$) measurements of these two compounds and the results are shown in Fig.~\ref{fig3}. Figure~\ref{fig3}(a) shows the specific heat results of \tszo down to 70~mK. There is no sharp $\lambda$-type peak expected for a well-defined phase transition. Instead, there is a kink around 7~K. This kink temperature almost coincides with the temperature when the susceptibility deviates from the Curie-Weiss behavior shown in Fig.~\ref{fig1}(e). In some other QSL candidates, there is a more obvious hump which probably corresponds to the establishment of short-range spin correlations~\cite{prl98_107204,np4_459,sr5_16419,PhysRevLett.120.087201}. We conjecture that the underlying physics for the kink observed here is similar. Another possible explanation is that it results from the thermal population of the low-lying CEF level located at 1.6~meV as shown in Fig.~\ref{fig2}(a). As for a two-level system with the energy splitting of 1.6~meV, it gives rise to a maximum at 7.7~K which is close to the kink temperature observed in both $\chi(T)$ and $C(T)$. Generally speaking, the specific heat at low temperatures can be fitted as $C_{\rm p} = \gamma T + \beta T^3$ for a system with gapless fermionic excitations~(Refs.~\onlinecite{np4_459,nc2_275}), where the linear $T$ term and $T^3$ term denote electronic and phononic contributions, respectively. Compared with the nonmagnetic reference compound \lszo behaving well as $C_{\rm p}/T \sim T^2$, which is reasonable, since it is an insulator~\cite{PhysRevB.98.174404}, \tszo has a large residual linear term. In the inset of Fig.~\ref{fig3}(a), we plot $C_{\rm p}/T$ as a function of $T^2$ to focus on the low-temperature part, and the linear extrapolation to absolute zero temperature yields a finite linear term coefficient $\gamma \sim$ 31.5(6)~mJ~mol$^{-1}$~K$^{-2}$. Such an observation is quite unusual for an insulator, and is often interpreted to be due to the fermionic fractional excitations such as spinons of a QSL~\cite{np4_459,nc2_275,sr5_16419}. The specific heat results of \tsmo depicted in Fig.~\ref{fig3}(b) exhibit similar behaviors. The kink shifts to a higher temperature of around 17~K, and it may reflect the fact that the exchange interaction is stronger than that in \tszo, which is consistent with the higher $\Theta_{\rm CW}$ in \tsmo. We believe the kink should have the same origin as discussed earlier for \tszo. The specific heat also exhibits a linear term, as shown in the inset of Fig.~\ref{fig3}(b). However, the $\gamma$ value of 8.9(7)~mJ~mol$^{-1}$~K$^{-2}$ is more than three times smaller than that of 31.5(6)~mJ~mol$^{-1}$~K$^{-2}$ of \tszo, indicating much less density of states  at low energies in \tsmo. Considering the substantial amount of disorder in \tszo, we believe that the enhancement of the $\gamma$ is due to the disorder resulting from the strong site mixing of Tm$^{3+}$ and Zn$^{2+}$ ions. We also present the magnetic specific heat~($C_{\rm m}$) results of these two compounds in Figs.~\ref{fig3}(c) and~\ref{fig3}(d). The kinks around 7~K in \tszo and 17~K in \tsmo are more clearly shown after subtracting the phonon contributions.

\subsection{INS spectra}

\begin{figure*}[htb]
  \centering
  \includegraphics[width=0.9\linewidth]{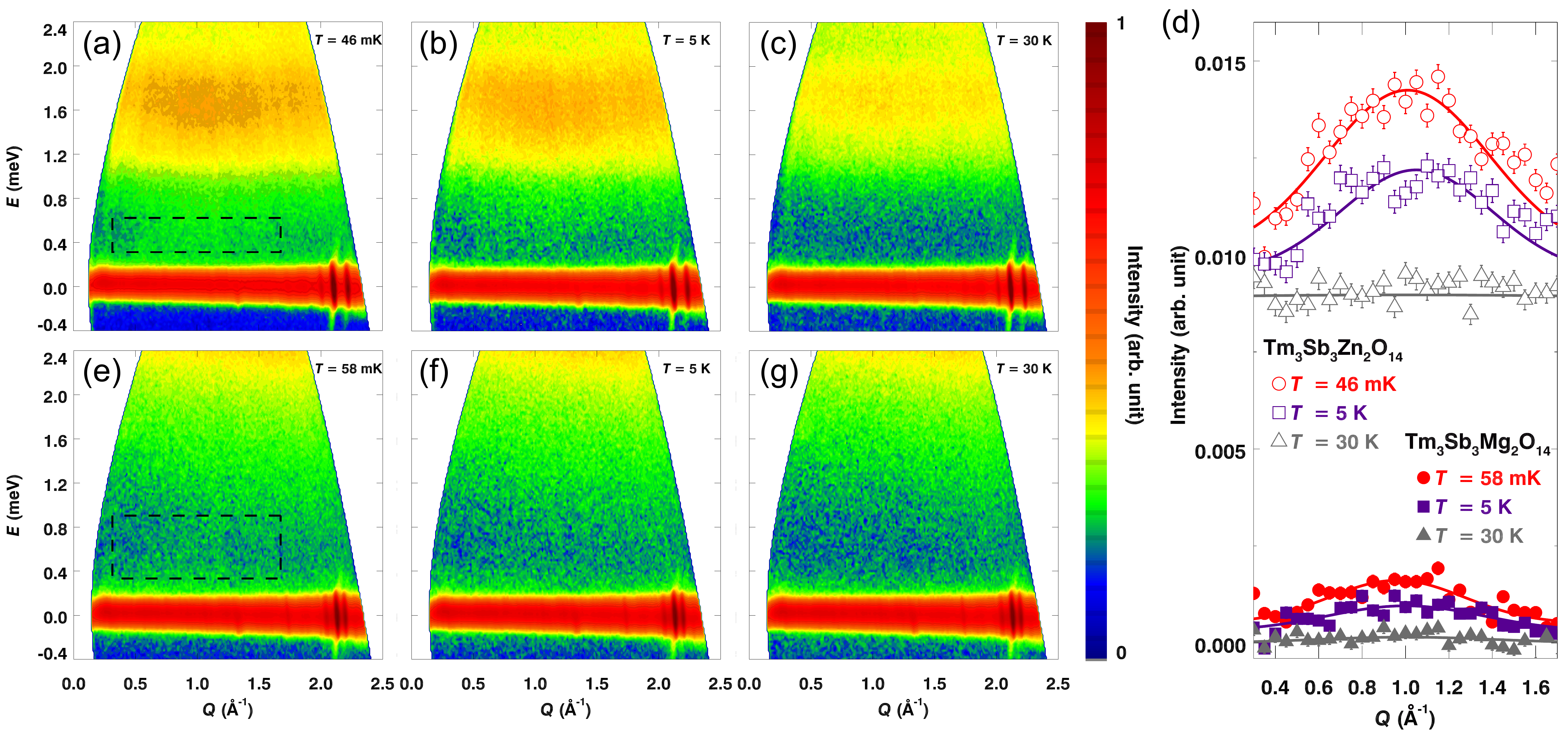}
  \caption{Inelastic neutron scattering spectra of \tszo (a)-(c) and \tsmo (e)-(g) measured on PELICAN spectrometer~\cite{doi:10.7566/JPSJS.82SA.SA027}. (d) shows the wave-vector $\bm{Q}$-dependence of the energy-integrated intensity of \tszo and \tsmo. The energy and $\bm{Q}$ range are marked with the dashed rectangular in (a) and (e). Because the signals for \tsmo are too weak, to visualize the $\bm{Q}$-dependence of the intensity in (d), we used the 60-K data as the background and subtracted it. Solid lines are guides to the eye. }
  \label{fig4}
\end{figure*}

\begin{figure}[htb]
\centerline{\includegraphics[width=3.4in]{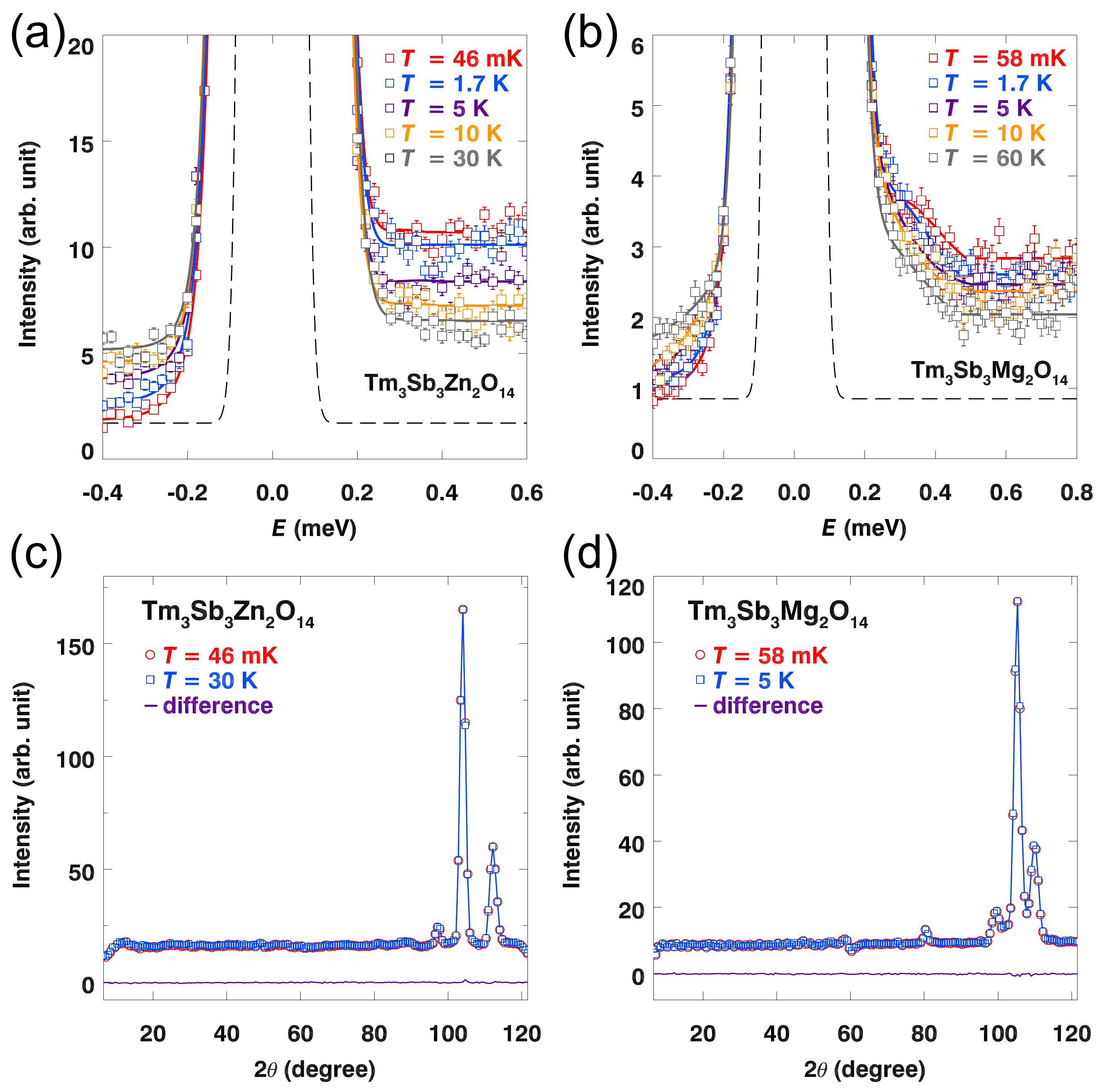}}
\caption{Energy dependence of the integrated neutron scattering intensity in the $\bm{Q}$ range between 0.6 to 1.4~\AA$^{-1}$ for \tszo (a) and \tsmo (b). Solid lines are guides to the eye. Dashed lines represent the instrumental resolutions. Elastic neutron scattering data for (c) \tszo and (d) \tsmo, obtained by integrating the intensity in an energy window of [-0.1,0.1]~meV. The data were collected on a time-of-flight spectrometer PELICAN at various temperatures.
\label{fig5}}
\end{figure}

We now turn to the INS measurements at low energies to gain further insights into the magnetic state. Figures~\ref{fig4}(a)-\ref{fig4}(c) show the magnetic excitation spectra (raw data) collected for the polycrystalline sample of \tszo at three characteristic temperatures. The INS results contain two pronounced features. First is the flat excitation band between $E= 1.1$ and 2.1~meV. These excitations are the low-lying CEF excitations of the Tm$^{3+}$ ions as also shown in Fig.~\ref{fig2}(a). Second is another broad and nearly flat excitation band below the CEF exciations, approximately in the energy range of 0.3 to 0.6~meV. The intensity weakens at 5~K [Fig.~\ref{fig4}(b)] and disappears at 30~K [Fig.~\ref{fig4}(c)]. To investigate the low-energy excitations in detail, we integrate the intensities with energy ranging from 0.3 to 0.6~meV to avoid contaminations from the CEF excitations and elastic scattering, and plot the integrated intensity as a function of $\bm{Q}$ in Fig.~\ref{fig4}(d). It is clear that there is a broad peak centered at $\bm{Q}~\sim$~1~\AA$^{-1}$ which corresponds to the $\Gamma$ point in the second Brillouin zone of a kagom\'{e} lattice. The $\bm{Q}$ and temperature dependence of these excitations indicate that they are of magnetic origin~\cite{PhysRevLett.120.087201,np15.262}. Figures~\ref{fig4}(e)-\ref{fig4}(g) show the INS spectra of \tsmo. As discussed earlier and shown in Fig.~\ref{fig2}(b), the first CEF level of \tsmo is 4~meV, which is beyond the energy range in Fig.~\ref{fig4}. As a result, we do not observe the CEF excitations in Figs.~\ref{fig4}(e)-\ref{fig4}(g) for \tsmo. The INS measurements were performed on the samples with similar weights and thus similar amount of magnetic Tm$^{3+}$ ions. Furthermore, we used the same experimental setup and equal counting time. Therefore, Figs.~\ref{fig4}(a)-~\ref{fig4}(c) and~\ref{fig4}(e)-\ref{fig4}(g) can be compared directly. Different from that in \tszo, the low-energy excitations are barely visible in Fig.~\ref{fig4}(e) for \tsmo. We integrate the intensity between 0.3 and 0.9~meV, but since the intensity is too weak, we need to subtract the background data at 60~K to make it visible. The so-obtained intensity as a function of $\bm{Q}$ is shown in Fig.~\ref{fig4}(d). There also appears to be a broad peak centered at $\bm{Q}~\sim$~1~\AA$^{-1}$, and the intensity increases as the temperature decreases. However, the overall intensities are significantly weaker than those in \tszo.

To further investigate the broad excitations centered at $\bm{Q}~\sim$~1~\AA$^{-1}$, we integrated the intensities with $\bm{Q}$ ranging from 0.6 and 1.4~\AA$^{-1}$ at various temperatures, and the integrated intensities are plotted as a function of energy in Fig.~\ref{fig5}. The intensities follow similar temperature dependence for both compounds: on the energy-loss side ($E>0$), intensities increase with decreasing temperature; on the energy-gain ($E<0$) side, the intensities are suppressed at low temperatures due to the detail balance. These results indicate that the intensities are resulting from intrinsic magnetic excitations. Moreover, the magnetic signals are dominated by the inelastic scattering, while the quasi-elastic scattering has no temperature dependence. It is reasonable that the spectral weight is mainly distributed in the inelastic channels for a system with moderate spin interactions in which $\Theta_{\rm {CW}}$ is -17.4 and -28.7~K for \tszo and \tsmo, respectively. We integrate the elastic channels and plot the data collected at $\sim$50~mK, and there is no magnetic Bragg peak observed in Figs.~\ref{fig5}(c) and~\ref{fig5}(d), which also proves the absence of magnetic order for both compounds down to $\sim$50~mK.

\section{Discussions and Conclusions}

How do we understand the INS results given the strong site mixing demonstrated from the XRD, CEF, and specific heat results? At first glance, the low-energy magnetic excitations in \tszo are distinct from conventional spin-wave excitations, but similar to the spectrum expected from the deconfined spinons in QSLs~\cite{nature492_406,nature540_559,np12_942,np15.262,np15_1052}. However, we believe the disorder-induced low-energy excitations will be a more natural explanation. Since the low-lying state of Tm$^{3+}$ at the original sites is a nearly degenerate non-Kramers doublet and dominated by $J_z=\pm6$ components by our CEF analysis, the effective spin residing in the doublet is more likely to behave as a multipole, which is not directly accessible for neutron scattering that has a selection rule of $\Delta S = \pm1$~(Ref.~\onlinecite{RevModPhys.85.367}). This explains nicely why the low-energy excitations are so weak in \tsmo, which has much less disorder. We need to point out that a well-isolated singlet ground state of Tm$^{3+}$ at the original sites in \tsmo was also proposed\cite{dun2020effective}. Although we cannot rule out this possibility, it will not affect the conclusion for the absence of low-energy magnetic excitations resulting from the original sites, as there will be no extra magnetic excitations below the first CEF level from the isolated singlet ground state. We believe the spin excitations resulting from the effective spin of Tm$^{3+}$ in the Zn$^{2+}$(2) sites will not give rise to the intensities observed in the INS experiment either as these two sites have similar CEF environments with eight O$^{2-}$ anions surrounding them~\cite{doi:10.1002/pssb.201600256}. For \tszo with a strong disorder effect, a large amount of Tm$^{3+}$ ions leave their original sites and occupy Zn$^{2+}$(1) position which is surrounded by six O$^{2-}$ anions, different from the situation in their original sites~\cite{doi:10.1002/pssb.201600256}. The CEF environment with a higher symmetry may preserve the effective spin-1/2 resulting from the dominating $J_z=\pm6$ and $\pm3$ components in the non-Kramers doublet, and in this case the dipole moments of the non-Kramers doublet are effective and can give rise to the low-energy magnetic excitations observed in the INS
experiment\cite{PhysRevX.10.011007}. Furthermore, the Zn$^{2+}$(1) sites are at the center of the hexagon. This will tune the kagom\'e lattice to triangular lattice effectively. In this case, the peak center of the excitation intensity $\bm{Q}\sim$1~\AA$^{-1}$ corresponds to the M point of the twice-enlarged Brillouin zone in the triangular lattice due to the reduction of the $a$ axis by half. This is fully consistent with the observation of negative Curie-Weiss temperature for the dominant antiferromagnetic interactions. Of course, INS measurements on single crystals are desirable in order to better reveal the momentum distribution of the spectral weight. Nevertheless, our observation of the site-mixing-induced broad gapless excitations is fully in line with previous reports in YbZnGaO$_4$~(Ref.~\onlinecite{PhysRevLett.120.087201}), Yb$_2$TiO$_5$~(Ref.~\onlinecite{PhysRevB.96.174418}), and $\kappa$-(ET)$_2$Cu[N(CN)$_2$]Cl~(Ref.~\onlinecite{PhysRevLett.115.077001}).

At this point, we conclude that the disorder resulting from the site mixing of the magnetic Tm$^{3+}$ ions and the nonmagnetic ones is responsible for the absence of magnetic order, CEF excitations, specific heat, and the INS spectra. Intriguingly, the amount of disorder is manifested in the value of the residual linear term in the specific heat and the strength of the low-energy magnetic excitations. Then a natural question is what is the ground state with no disorder? One possibility is the valence-bond-solid state, in which two nearest-neighbor antiparallel spins form a singlet---such a state will not exhibit a magnetic order either~\cite{nc_Itamar,PhysRevX.8.031028,PhysRevX.8.041040,nc10_2561,PhysRevB.101.140401}. Another possibility is the gapped QSL state. In both cases, disorder will induce the low-energy magnetic excitations that contribute to the specific heat and magnetic neutron scattering. Although to distinguish these two states requires further experimental efforts, our results here demonstrate directly that by bringing disorder into such a geometrically frustrated system, a spin-liquid-like state can be induced. We believe this conclusion holds for a broad class of frustrated magnets including both the organic and inorganic compounds in the presence of either magnetic or nonmagnetic disorder. With many efforts, both experimental~\cite{np13_117,PhysRevLett.118.107202,PhysRevB.94.060409,PhysRevLett.120.087201,PhysRevX.8.031001,PhysRevB.96.174418,PhysRevLett.115.077001} and theoretical~\cite{PhysRevLett.119.157201,PhysRevB.97.184413,PhysRevLett.118.087203,PhysRevLett.120.207203,nc_Itamar,PhysRevX.8.031028,PhysRevX.8.041040,PhysRevLett.123.087201}, the role of disorder in the realization towards the spin-liquid-like behaviors has been made prominent. This also poses a great challenge for experimentalists in identifying a QSL~\cite{Wen2019Experimental}.

\section{Acknowledgements}

The work was supported by the National Natural Science Foundation of China with Grant Nos~11822405, 12074174, 11674157, 11674158, 11774152, 11904170, 12004249, and 12004251, National Key
Projects for Research and Development of China with Grant No.~2016YFA0300401, Natural Science Foundation of Jiangsu Province with Grant Nos~BK20180006 and BK20190436, Shanghai Sailing Program with Grant No.~20YF1430600, Fundamental Research Funds for the Central Universities with Grant No.~020414380117, and the Office of International Cooperation and Exchanges of Nanjing University. Hai-Feng Li acknowledges financial support from the Science and Technology Development Fund, Macau SAR (File Nos~028/2017/A1 and 0051/2019/AFJ). We acknowledge stimulating discussions with Itamar Kimchi. We thank the ACNS at ANSTO for the access of neutron-scattering facilities through Proposals 7312 and 7314, and the excellent support from Gene Davidson from the Sample Environment Group for setting up the dilution refrigerator on PELICAN.

%\bibliography{qsl,topo}

\begin{thebibliography}{70}%
\makeatletter
\providecommand \@ifxundefined [1]{%
 \@ifx{#1\undefined}
}%
\providecommand \@ifnum [1]{%
 \ifnum #1\expandafter \@firstoftwo
 \else \expandafter \@secondoftwo
 \fi
}%
\providecommand \@ifx [1]{%
 \ifx #1\expandafter \@firstoftwo
 \else \expandafter \@secondoftwo
 \fi
}%
\providecommand \natexlab [1]{#1}%
\providecommand \enquote  [1]{``#1''}%
\providecommand \bibnamefont  [1]{#1}%
\providecommand \bibfnamefont [1]{#1}%
\providecommand \citenamefont [1]{#1}%
\providecommand \href@noop [0]{\@secondoftwo}%
\providecommand \href [0]{\begingroup \@sanitize@url \@href}%
\providecommand \@href[1]{\@@startlink{#1}\@@href}%
\providecommand \@@href[1]{\endgroup#1\@@endlink}%
\providecommand \@sanitize@url [0]{\catcode `\\12\catcode `\$12\catcode
  `\&12\catcode `\#12\catcode `\^12\catcode `\_12\catcode `\%12\relax}%
\providecommand \@@startlink[1]{}%
\providecommand \@@endlink[0]{}%
\providecommand \url  [0]{\begingroup\@sanitize@url \@url }%
\providecommand \@url [1]{\endgroup\@href {#1}{\urlprefix }}%
\providecommand \urlprefix  [0]{URL }%
\providecommand \Eprint [0]{\href }%
\providecommand \doibase [0]{http://dx.doi.org/}%
\providecommand \selectlanguage [0]{\@gobble}%
\providecommand \bibinfo  [0]{\@secondoftwo}%
\providecommand \bibfield  [0]{\@secondoftwo}%
\providecommand \translation [1]{[#1]}%
\providecommand \BibitemOpen [0]{}%
\providecommand \bibitemStop [0]{}%
\providecommand \bibitemNoStop [0]{.\EOS\space}%
\providecommand \EOS [0]{\spacefactor3000\relax}%
\providecommand \BibitemShut  [1]{\csname bibitem#1\endcsname}%
\let\auto@bib@innerbib\@empty
%</preamble>
\bibitem [{\citenamefont {Anderson}(1973)}]{Anderson1973153}%
  \BibitemOpen
  \bibfield  {author} {\bibinfo {author} {\bibfnamefont {P.W.}\ \bibnamefont
  {Anderson}},\ }\bibfield  {title} {\enquote {\bibinfo {title} {{Resonating
  valence bonds: A new kind of insulator?}}}\ }\href@noop {} {\bibfield
  {journal} {\bibinfo  {journal} {Mater. Res. Bull.}\ }\textbf {\bibinfo
  {volume} {8}},\ \bibinfo {pages} {153} (\bibinfo {year}
  {1973})}\BibitemShut {NoStop}%
\bibitem [{\citenamefont {Balents}(2010)}]{nature464_199}%
  \BibitemOpen
  \bibfield  {author} {\bibinfo {author} {\bibfnamefont {Leon}\ \bibnamefont
  {Balents}},\ }\bibfield  {title} {\enquote {\bibinfo {title} {Spin liquids in
  frustrated magnets},}\ }\href@noop {} {\bibfield  {journal} {\bibinfo
  {journal} {Nature}\ }\textbf {\bibinfo {volume} {464}},\ \bibinfo {pages}
  {199} (\bibinfo {year} {2010})}\BibitemShut {NoStop}%
\bibitem [{\citenamefont {Imai}\ and\ \citenamefont
  {Lee}(2016)}]{imai2016quantum}%
  \BibitemOpen
  \bibfield  {author} {\bibinfo {author} {\bibfnamefont {Takashi}\ \bibnamefont
  {Imai}}\ and\ \bibinfo {author} {\bibfnamefont {Young~S.}\ \bibnamefont
  {Lee}},\ }\bibfield  {title} {\enquote {\bibinfo {title} {Do quantum spin
  liquids exist?}}\ }\href@noop {} {\bibfield  {journal} {\bibinfo  {journal}
  {Phys. Today}\ }\textbf {\bibinfo {volume} {69}},\ \bibinfo {pages}
  {30} (\bibinfo {year} {2016})}\BibitemShut {NoStop}%
\bibitem [{\citenamefont {Kitaev}(2003)}]{Kitaev20032}%
  \BibitemOpen
  \bibfield  {author} {\bibinfo {author} {\bibfnamefont {A.~Yu.}\ \bibnamefont
  {Kitaev}},\ }\bibfield  {title} {\enquote {\bibinfo {title} {Fault-tolerant
  quantum computation by anyons},}\ }\href {\doibase
  http://dx.doi.org/10.1016/S0003-4916(02)00018-0} {\bibfield  {journal}
  {\bibinfo  {journal} {Ann. Phys.}\ }\textbf {\bibinfo {volume} {303}},\
  \bibinfo {pages} {2} (\bibinfo {year} {2003})}\BibitemShut {NoStop}%
\bibitem [{\citenamefont {Barkeshli}\ \emph {et~al.}(2014)\citenamefont
  {Barkeshli}, \citenamefont {Berg},\ and\ \citenamefont
  {Kivelson}}]{Barkeshli722}%
  \BibitemOpen
  \bibfield  {author} {\bibinfo {author} {\bibfnamefont {Maissam}\ \bibnamefont
  {Barkeshli}}, \bibinfo {author} {\bibfnamefont {Erez}\ \bibnamefont {Berg}},
  \ and\ \bibinfo {author} {\bibfnamefont {Steven}\ \bibnamefont {Kivelson}},\
  }\bibfield  {title} {\enquote {\bibinfo {title} {Coherent transmutation of
  electrons into fractionalized anyons},}\ }\href {\doibase
  10.1126/science.1253251} {\bibfield  {journal} {\bibinfo  {journal}
  {Science}\ }\textbf {\bibinfo {volume} {346}},\ \bibinfo {pages} {722}
  (\bibinfo {year} {2014})}\BibitemShut
  {NoStop}%
\bibitem [{\citenamefont {Anderson}(1987)}]{anderson1}%
  \BibitemOpen
  \bibfield  {author} {\bibinfo {author} {\bibfnamefont {P.~W.}\ \bibnamefont
  {Anderson}},\ }\bibfield  {title} {\enquote {\bibinfo {title} {{The
  Resonating Valence Bond State in La$_2$CuO$_4$ and Superconductivity}},}\
  }\href@noop {} {\bibfield  {journal} {\bibinfo  {journal} {Science}\ }\textbf
  {\bibinfo {volume} {235}},\ \bibinfo {pages} {1196} (\bibinfo {year}
  {1987})}\BibitemShut {NoStop}%
\bibitem [{\citenamefont {Baskaran}\ \emph {et~al.}(1987)\citenamefont
  {Baskaran}, \citenamefont {Zou},\ and\ \citenamefont
  {Anderson}}]{Baskaran1987973}%
  \BibitemOpen
  \bibfield  {author} {\bibinfo {author} {\bibfnamefont {G.}~\bibnamefont
  {Baskaran}}, \bibinfo {author} {\bibfnamefont {Z.}~\bibnamefont {Zou}}, \
  and\ \bibinfo {author} {\bibfnamefont {P.W.}\ \bibnamefont {Anderson}},\
  }\bibfield  {title} {\enquote {\bibinfo {title} {{The resonating valence bond
  state and high-Tc superconductivity --- A mean field theory}},}\ }\href
  {\doibase http://dx.doi.org/10.1016/0038-1098(87)90642-9} {\bibfield
  {journal} {\bibinfo  {journal} {Solid State Commun.}\ }\textbf
  {\bibinfo {volume} {63}},\ \bibinfo {pages} {973} (\bibinfo {year}
  {1987})}\BibitemShut {NoStop}%
\bibitem [{\citenamefont {Lee}\ \emph {et~al.}(2006)\citenamefont {Lee},
  \citenamefont {Nagaosa},\ and\ \citenamefont {Wen}}]{lee:17}%
  \BibitemOpen
  \bibfield  {author} {\bibinfo {author} {\bibfnamefont {Patrick~A.}\
  \bibnamefont {Lee}}, \bibinfo {author} {\bibfnamefont {Naoto}\ \bibnamefont
  {Nagaosa}}, \ and\ \bibinfo {author} {\bibfnamefont {Xiao-Gang}\ \bibnamefont
  {Wen}},\ }\bibfield  {title} {\enquote {\bibinfo {title} {{Doping a Mott
  insulator: Physics of high-temperature superconductivity}},}\ }\href@noop {}
  {\bibfield  {journal} {\bibinfo  {journal} {Rev. Mod. Phys.}\ }\textbf
  {\bibinfo {volume} {78}},\ \bibinfo {pages} {17} (\bibinfo {year}
  {2006})}\BibitemShut {NoStop}%
\bibitem [{\citenamefont {N\'{e}el}(1971)}]{Neel1985}%
  \BibitemOpen
  \bibfield  {author} {\bibinfo {author} {\bibfnamefont {Louis}\ \bibnamefont
  {N\'{e}el}},\ }\bibfield  {title} {\enquote {\bibinfo {title} {Magnetism and
  local molecular field},}\ }\href {\doibase 10.1126/science.174.4013.985}
  {\bibfield  {journal} {\bibinfo  {journal} {Science}\ }\textbf {\bibinfo
  {volume} {174}},\ \bibinfo {pages} {985} (\bibinfo {year} {1971})}\BibitemShut
  {NoStop}%
\bibitem [{\citenamefont {Ramirez}(1994)}]{arms24_453}%
  \BibitemOpen
  \bibfield  {author} {\bibinfo {author} {\bibfnamefont {A.~P.}\ \bibnamefont
  {Ramirez}},\ }\bibfield  {title} {\enquote {\bibinfo {title} {Strongly
  geometrically frustrated magnets},}\ }\href {\doibase
  10.1146/annurev.ms.24.080194.002321} {\bibfield  {journal} {\bibinfo
  {journal} {Annu. Rev. Mater. Sci.}\ }\textbf {\bibinfo {volume}
  {24}},\ \bibinfo {pages} {453} (\bibinfo {year} {1994})}\BibitemShut
  {NoStop}%
\bibitem [{\citenamefont {Savary}\ and\ \citenamefont
  {Balents}(2017{\natexlab{a}})}]{0034-4885-80-1-016502}%
  \BibitemOpen
  \bibfield  {author} {\bibinfo {author} {\bibfnamefont {Lucile}\ \bibnamefont
  {Savary}}\ and\ \bibinfo {author} {\bibfnamefont {Leon}\ \bibnamefont
  {Balents}},\ }\bibfield  {title} {\enquote {\bibinfo {title} {Quantum spin
  liquids: a review},}\ }\href
  {http://stacks.iop.org/0034-4885/80/i=1/a=016502} {\bibfield  {journal}
  {\bibinfo  {journal} {Rep. Prog. Phys.}\ }\textbf {\bibinfo
  {volume} {80}},\ \bibinfo {pages} {016502} (\bibinfo {year}
  {2017}{\natexlab{a}})}\BibitemShut {NoStop}%
\bibitem [{\citenamefont {Zhou}\ \emph {et~al.}(2017)\citenamefont {Zhou},
  \citenamefont {Kanoda},\ and\ \citenamefont {Ng}}]{RevModPhys.89.025003}%
  \BibitemOpen
  \bibfield  {author} {\bibinfo {author} {\bibfnamefont {Yi}~\bibnamefont
  {Zhou}}, \bibinfo {author} {\bibfnamefont {Kazushi}\ \bibnamefont {Kanoda}},
  \ and\ \bibinfo {author} {\bibfnamefont {Tai-Kai}\ \bibnamefont {Ng}},\
  }\bibfield  {title} {\enquote {\bibinfo {title} {Quantum spin liquid
  states},}\ }\href {\doibase 10.1103/RevModPhys.89.025003} {\bibfield
  {journal} {\bibinfo  {journal} {Rev. Mod. Phys.}\ }\textbf {\bibinfo {volume}
  {89}},\ \bibinfo {pages} {025003} (\bibinfo {year} {2017})}\BibitemShut
  {NoStop}%
\bibitem [{\citenamefont {Wen}\ \emph {et~al.}(2019)\citenamefont {Wen},
  \citenamefont {Yu}, \citenamefont {Li}, \citenamefont {Yu},\ and\
  \citenamefont {Li}}]{Wen2019Experimental}%
  \BibitemOpen
  \bibfield  {author} {\bibinfo {author} {\bibfnamefont {Jinsheng}\
  \bibnamefont {Wen}}, \bibinfo {author} {\bibfnamefont {Shun-Li}\ \bibnamefont
  {Yu}}, \bibinfo {author} {\bibfnamefont {Shiyan}\ \bibnamefont {Li}},
  \bibinfo {author} {\bibfnamefont {Weiqiang}\ \bibnamefont {Yu}}, \ and\
  \bibinfo {author} {\bibfnamefont {Jian-Xin}\ \bibnamefont {Li}},\ }\bibfield
  {title} {\enquote {\bibinfo {title} {Experimental identification of quantum
  spin liquids},}\ }\href@noop {} {\bibfield  {journal} {\bibinfo  {journal}
  {npj Quant. Mater.}\ }\textbf {\bibinfo {volume} {4}},\ \bibinfo {pages}
  {12} (\bibinfo {year} {2019})}\BibitemShut {NoStop}%
\bibitem [{\citenamefont {Shimizu}\ \emph {et~al.}(2003)\citenamefont
  {Shimizu}, \citenamefont {Miyagawa}, \citenamefont {Kanoda}, \citenamefont
  {Maesato},\ and\ \citenamefont {Saito}}]{PhysRevLett.91.107001}%
  \BibitemOpen
  \bibfield  {author} {\bibinfo {author} {\bibfnamefont {Y.}~\bibnamefont
  {Shimizu}}, \bibinfo {author} {\bibfnamefont {K.}~\bibnamefont {Miyagawa}},
  \bibinfo {author} {\bibfnamefont {K.}~\bibnamefont {Kanoda}}, \bibinfo
  {author} {\bibfnamefont {M.}~\bibnamefont {Maesato}}, \ and\ \bibinfo
  {author} {\bibfnamefont {G.}~\bibnamefont {Saito}},\ }\bibfield  {title}
  {\enquote {\bibinfo {title} {{Spin Liquid State in an Organic Mott Insulator
  with a Triangular Lattice}},}\ }\href {\doibase
  10.1103/PhysRevLett.91.107001} {\bibfield  {journal} {\bibinfo  {journal}
  {Phys. Rev. Lett.}\ }\textbf {\bibinfo {volume} {91}},\ \bibinfo {pages}
  {107001} (\bibinfo {year} {2003})}\BibitemShut {NoStop}%
\bibitem [{\citenamefont {Kurosaki}\ \emph {et~al.}(2005)\citenamefont
  {Kurosaki}, \citenamefont {Shimizu}, \citenamefont {Miyagawa}, \citenamefont
  {Kanoda},\ and\ \citenamefont {Saito}}]{PhysRevLett.95.177001}%
  \BibitemOpen
  \bibfield  {author} {\bibinfo {author} {\bibfnamefont {Y.}~\bibnamefont
  {Kurosaki}}, \bibinfo {author} {\bibfnamefont {Y.}~\bibnamefont {Shimizu}},
  \bibinfo {author} {\bibfnamefont {K.}~\bibnamefont {Miyagawa}}, \bibinfo
  {author} {\bibfnamefont {K.}~\bibnamefont {Kanoda}}, \ and\ \bibinfo {author}
  {\bibfnamefont {G.}~\bibnamefont {Saito}},\ }\bibfield  {title} {\enquote
  {\bibinfo {title} {{Mott Transition from a Spin Liquid to a Fermi Liquid in
  the Spin-Frustrated Organic Conductor $\kappa$-(ET)$_2$Cu$_2$(CN)$_3$}},}\
  }\href {\doibase 10.1103/PhysRevLett.95.177001} {\bibfield  {journal}
  {\bibinfo  {journal} {Phys. Rev. Lett.}\ }\textbf {\bibinfo {volume} {95}},\
  \bibinfo {pages} {177001} (\bibinfo {year} {2005})}\BibitemShut {NoStop}%
\bibitem [{\citenamefont {Ohira}\ \emph {et~al.}(2006)\citenamefont {Ohira},
  \citenamefont {Shimizu}, \citenamefont {Kanoda},\ and\ \citenamefont
  {Saito}}]{Ohira2006}%
  \BibitemOpen
  \bibfield  {author} {\bibinfo {author} {\bibfnamefont {S.}~\bibnamefont
  {Ohira}}, \bibinfo {author} {\bibfnamefont {Y.}~\bibnamefont {Shimizu}},
  \bibinfo {author} {\bibfnamefont {K.}~\bibnamefont {Kanoda}}, \ and\ \bibinfo
  {author} {\bibfnamefont {G.}~\bibnamefont {Saito}},\ }\bibfield  {title}
  {\enquote {\bibinfo {title} {{Spin liquid state in
  $\kappa$-(BEDT-TTF)$_2$Cu$_2$(CN)$_3$ studied by muon spin relaxation
  method}},}\ }\href {\doibase 10.1007/BF02679485} {\bibfield  {journal}
  {\bibinfo  {journal} {J. Low Temp. Phys.}\ }\textbf {\bibinfo
  {volume} {142}},\ \bibinfo {pages} {153} (\bibinfo {year}
  {2006})}\BibitemShut {NoStop}%
\bibitem [{\citenamefont {Yamashita}\ \emph {et~al.}(2008)\citenamefont
  {Yamashita}, \citenamefont {Nakazawa}, \citenamefont {Oguni}, \citenamefont
  {Oshima}, \citenamefont {Nojiri}, \citenamefont {Shimizu}, \citenamefont
  {Miyagawa},\ and\ \citenamefont {Kanoda}}]{np4_459}%
  \BibitemOpen
  \bibfield  {author} {\bibinfo {author} {\bibfnamefont {Satoshi}\ \bibnamefont
  {Yamashita}}, \bibinfo {author} {\bibfnamefont {Yasuhiro}\ \bibnamefont
  {Nakazawa}}, \bibinfo {author} {\bibfnamefont {Masaharu}\ \bibnamefont
  {Oguni}}, \bibinfo {author} {\bibfnamefont {Yugo}\ \bibnamefont {Oshima}},
  \bibinfo {author} {\bibfnamefont {Hiroyuki}\ \bibnamefont {Nojiri}}, \bibinfo
  {author} {\bibfnamefont {Yasuhiro}\ \bibnamefont {Shimizu}}, \bibinfo
  {author} {\bibfnamefont {Kazuya}\ \bibnamefont {Miyagawa}}, \ and\ \bibinfo
  {author} {\bibfnamefont {Kazushi}\ \bibnamefont {Kanoda}},\ }\bibfield
  {title} {\enquote {\bibinfo {title} {Thermodynamic properties of a spin-1/2
  spin-liquid state in a $\kappa$-type organic salt},}\ }\href@noop {}
  {\bibfield  {journal} {\bibinfo  {journal} {Nat. Phys.}\ }\textbf {\bibinfo
  {volume} {4}},\ \bibinfo {pages} {459} (\bibinfo {year}
  {2008})}\BibitemShut {NoStop}%
\bibitem [{\citenamefont {Yamashita}\ \emph {et~al.}(2009)\citenamefont
  {Yamashita}, \citenamefont {Nakata}, \citenamefont {Kasahara}, \citenamefont
  {Sasaki}, \citenamefont {Yoneyama}, \citenamefont {Kobayashi}, \citenamefont
  {Fujimoto}, \citenamefont {Shibauchi},\ and\ \citenamefont
  {Matsuda}}]{np5_44}%
  \BibitemOpen
  \bibfield  {author} {\bibinfo {author} {\bibfnamefont {Minoru}\ \bibnamefont
  {Yamashita}}, \bibinfo {author} {\bibfnamefont {Norihito}\ \bibnamefont
  {Nakata}}, \bibinfo {author} {\bibfnamefont {Yuichi}\ \bibnamefont
  {Kasahara}}, \bibinfo {author} {\bibfnamefont {Takahiko}\ \bibnamefont
  {Sasaki}}, \bibinfo {author} {\bibfnamefont {Naoki}\ \bibnamefont
  {Yoneyama}}, \bibinfo {author} {\bibfnamefont {Norio}\ \bibnamefont
  {Kobayashi}}, \bibinfo {author} {\bibfnamefont {Satoshi}\ \bibnamefont
  {Fujimoto}}, \bibinfo {author} {\bibfnamefont {Takasada}\ \bibnamefont
  {Shibauchi}}, \ and\ \bibinfo {author} {\bibfnamefont {Yuji}\ \bibnamefont
  {Matsuda}},\ }\bibfield  {title} {\enquote {\bibinfo {title}
  {{Thermal-transport measurements in a quantum spin-liquid state of the
  frustrated triangular magnet $\kappa$-(BEDT-TTF)$_2$Cu$_2$(CN)$_3$}},}\
  }\href@noop {} {\bibfield  {journal} {\bibinfo  {journal} {Nat. Phys.}\
  }\textbf {\bibinfo {volume} {5}},\ \bibinfo {pages} {44} (\bibinfo {year}
  {2009})}\BibitemShut {NoStop}%
\bibitem [{\citenamefont {Furukawa}\ \emph {et~al.}(2018)\citenamefont
  {Furukawa}, \citenamefont {Kobashi}, \citenamefont {Kurosaki}, \citenamefont
  {Miyagawa},\ and\ \citenamefont {Kanoda}}]{Furukawa2018Quasi}%
  \BibitemOpen
  \bibfield  {author} {\bibinfo {author} {\bibfnamefont {T.}~\bibnamefont
  {Furukawa}}, \bibinfo {author} {\bibfnamefont {K.}~\bibnamefont {Kobashi}},
  \bibinfo {author} {\bibfnamefont {Y.}~\bibnamefont {Kurosaki}}, \bibinfo
  {author} {\bibfnamefont {K.}~\bibnamefont {Miyagawa}}, \ and\ \bibinfo
  {author} {\bibfnamefont {K.}~\bibnamefont {Kanoda}},\ }\bibfield  {title}
  {\enquote {\bibinfo {title} {{Quasi-continuous transition from a Fermi liquid
  to a spin liquid in $\kappa$-(ET)$_2$Cu$_2$(CN)$_3$}},}\ }\href@noop {}
  {\bibfield  {journal} {\bibinfo  {journal} {Nat. Commun.}\ }\textbf
  {\bibinfo {volume} {9}},\ \bibinfo {pages} {307} (\bibinfo {year}
  {2018})}\BibitemShut {NoStop}%
\bibitem [{\citenamefont {Itou}\ \emph {et~al.}(2010)\citenamefont {Itou},
  \citenamefont {Oyamada}, \citenamefont {Maegawa},\ and\ \citenamefont
  {Kato}}]{np6_673}%
  \BibitemOpen
  \bibfield  {author} {\bibinfo {author} {\bibfnamefont {T.}~\bibnamefont
  {Itou}}, \bibinfo {author} {\bibfnamefont {A.}~\bibnamefont {Oyamada}},
  \bibinfo {author} {\bibfnamefont {S.}~\bibnamefont {Maegawa}}, \ and\
  \bibinfo {author} {\bibfnamefont {R.}~\bibnamefont {Kato}},\ }\bibfield
  {title} {\enquote {\bibinfo {title} {Instability of a quantum spin liquid in
  an organic triangular-lattice antiferromagnet},}\ }\href@noop {} {\bibfield
  {journal} {\bibinfo  {journal} {Nat. Phys.}\ }\textbf {\bibinfo {volume}
  {6}},\ \bibinfo {pages} {673} (\bibinfo {year} {2010})}\BibitemShut
  {NoStop}%
\bibitem [{\citenamefont {Yamashita}\ \emph {et~al.}(2010)\citenamefont
  {Yamashita}, \citenamefont {Nakata}, \citenamefont {Senshu}, \citenamefont
  {Nagata}, \citenamefont {Yamamoto}, \citenamefont {Kato}, \citenamefont
  {Shibauchi},\ and\ \citenamefont {Matsuda}}]{Yamashita1246}%
  \BibitemOpen
  \bibfield  {author} {\bibinfo {author} {\bibfnamefont {Minoru}\ \bibnamefont
  {Yamashita}}, \bibinfo {author} {\bibfnamefont {Norihito}\ \bibnamefont
  {Nakata}}, \bibinfo {author} {\bibfnamefont {Yoshinori}\ \bibnamefont
  {Senshu}}, \bibinfo {author} {\bibfnamefont {Masaki}\ \bibnamefont {Nagata}},
  \bibinfo {author} {\bibfnamefont {Hiroshi~M.}\ \bibnamefont {Yamamoto}},
  \bibinfo {author} {\bibfnamefont {Reizo}\ \bibnamefont {Kato}}, \bibinfo
  {author} {\bibfnamefont {Takasada}\ \bibnamefont {Shibauchi}}, \ and\
  \bibinfo {author} {\bibfnamefont {Yuji}\ \bibnamefont {Matsuda}},\ }\bibfield
   {title} {\enquote {\bibinfo {title} {Highly mobile gapless excitations in a
  two-dimensional candidate quantum spin liquid},}\ }\href {\doibase
  10.1126/science.1188200} {\bibfield  {journal} {\bibinfo  {journal}
  {Science}\ }\textbf {\bibinfo {volume} {328}},\ \bibinfo {pages} {1246}
  (\bibinfo {year} {2010})}\BibitemShut
  {NoStop}%
\bibitem [{\citenamefont {Yamashita}\ \emph {et~al.}(2011)\citenamefont
  {Yamashita}, \citenamefont {Yamamoto}, \citenamefont {Nakazawa},
  \citenamefont {Tamura},\ and\ \citenamefont {Kato}}]{nc2_275}%
  \BibitemOpen
  \bibfield  {author} {\bibinfo {author} {\bibfnamefont {Satoshi}\ \bibnamefont
  {Yamashita}}, \bibinfo {author} {\bibfnamefont {Takashi}\ \bibnamefont
  {Yamamoto}}, \bibinfo {author} {\bibfnamefont {Yasuhiro}\ \bibnamefont
  {Nakazawa}}, \bibinfo {author} {\bibfnamefont {Masafumi}\ \bibnamefont
  {Tamura}}, \ and\ \bibinfo {author} {\bibfnamefont {Reizo}\ \bibnamefont
  {Kato}},\ }\bibfield  {title} {\enquote {\bibinfo {title} {Gapless spin
  liquid of an organic triangular compound evidenced by thermodynamic
  measurements},}\ }\href@noop {} {\bibfield  {journal} {\bibinfo  {journal}
  {Nat. Commun.}\ }\textbf {\bibinfo {volume} {2}},\ \bibinfo {pages} {275}
  (\bibinfo {year} {2011})}\BibitemShut {NoStop}%
\bibitem [{\citenamefont {Helton}\ \emph {et~al.}(2007)\citenamefont {Helton},
  \citenamefont {Matan}, \citenamefont {Shores}, \citenamefont {Nytko},
  \citenamefont {Bartlett}, \citenamefont {Yoshida}, \citenamefont {Takano},
  \citenamefont {Suslov}, \citenamefont {Qiu}, \citenamefont {Chung},
  \citenamefont {Nocera},\ and\ \citenamefont {Lee}}]{prl98_107204}%
  \BibitemOpen
  \bibfield  {author} {\bibinfo {author} {\bibfnamefont {J.~S.}\ \bibnamefont
  {Helton}}, \bibinfo {author} {\bibfnamefont {K.}~\bibnamefont {Matan}},
  \bibinfo {author} {\bibfnamefont {M.~P.}\ \bibnamefont {Shores}}, \bibinfo
  {author} {\bibfnamefont {E.~A.}\ \bibnamefont {Nytko}}, \bibinfo {author}
  {\bibfnamefont {B.~M.}\ \bibnamefont {Bartlett}}, \bibinfo {author}
  {\bibfnamefont {Y.}~\bibnamefont {Yoshida}}, \bibinfo {author} {\bibfnamefont
  {Y.}~\bibnamefont {Takano}}, \bibinfo {author} {\bibfnamefont
  {A.}~\bibnamefont {Suslov}}, \bibinfo {author} {\bibfnamefont
  {Y.}~\bibnamefont {Qiu}}, \bibinfo {author} {\bibfnamefont {J.~H.}\
  \bibnamefont {Chung}}, \bibinfo {author} {\bibfnamefont {D.~G.}\ \bibnamefont
  {Nocera}}, \ and\ \bibinfo {author} {\bibfnamefont {Y.~S.}\ \bibnamefont
  {Lee}},\ }\bibfield  {title} {\enquote {\bibinfo {title} {{Spin Dynamics of
  the Spin-$1/2$ Kagome Lattice Antiferromagnet ZnCu$_3$(OH)$_6$Cl$_2$}},}\
  }\href@noop {} {\bibfield  {journal} {\bibinfo  {journal} {Phys. Rev. Lett.}\
  }\textbf {\bibinfo {volume} {98}},\ \bibinfo {pages} {107204} (\bibinfo
  {year} {2007})}\BibitemShut {NoStop}%
\bibitem [{\citenamefont {Han}\ \emph {et~al.}(2012)\citenamefont {Han},
  \citenamefont {Helton}, \citenamefont {Chu}, \citenamefont {Nocera},
  \citenamefont {Rodriguez-Rivera}, \citenamefont {Broholm},\ and\
  \citenamefont {Lee}}]{nature492_406}%
  \BibitemOpen
  \bibfield  {author} {\bibinfo {author} {\bibfnamefont {Tian-Heng}\
  \bibnamefont {Han}}, \bibinfo {author} {\bibfnamefont {Joel~S.}\ \bibnamefont
  {Helton}}, \bibinfo {author} {\bibfnamefont {Shaoyan}\ \bibnamefont {Chu}},
  \bibinfo {author} {\bibfnamefont {Daniel~G.}\ \bibnamefont {Nocera}},
  \bibinfo {author} {\bibfnamefont {Jose~A.}\ \bibnamefont {Rodriguez-Rivera}},
  \bibinfo {author} {\bibfnamefont {Collin}\ \bibnamefont {Broholm}}, \ and\
  \bibinfo {author} {\bibfnamefont {Young~S.}\ \bibnamefont {Lee}},\ }\bibfield
   {title} {\enquote {\bibinfo {title} {Fractionalized excitations in the
  spin-liquid state of a kagome-lattice antiferromagnet},}\ }\href@noop {}
  {\bibfield  {journal} {\bibinfo  {journal} {Nature}\ }\textbf {\bibinfo
  {volume} {492}},\ \bibinfo {pages} {406} (\bibinfo {year}
  {2012})}\BibitemShut {NoStop}%
\bibitem [{\citenamefont {Norman}(2016)}]{RevModPhys.88.041002}%
  \BibitemOpen
  \bibfield  {author} {\bibinfo {author} {\bibfnamefont {M.~R.}\ \bibnamefont
  {Norman}},\ }\bibfield  {title} {\enquote {\bibinfo {title}
  {\textit{Colloquium} : Herbertsmithite and the search for the quantum spin
  liquid},}\ }\href {\doibase 10.1103/RevModPhys.88.041002} {\bibfield
  {journal} {\bibinfo  {journal} {Rev. Mod. Phys.}\ }\textbf {\bibinfo {volume}
  {88}},\ \bibinfo {pages} {041002} (\bibinfo {year} {2016})}\BibitemShut
  {NoStop}%
\bibitem [{\citenamefont {Li}\ \emph {et~al.}(2015{\natexlab{a}})\citenamefont
  {Li}, \citenamefont {Liao}, \citenamefont {Zhang}, \citenamefont {Li},
  \citenamefont {Jin}, \citenamefont {Ling}, \citenamefont {Zhang},
  \citenamefont {Zou}, \citenamefont {Pi}, \citenamefont {Yang}, \citenamefont
  {Wang}, \citenamefont {Wu},\ and\ \citenamefont {Zhang}}]{sr5_16419}%
  \BibitemOpen
  \bibfield  {author} {\bibinfo {author} {\bibfnamefont {Yuesheng}\
  \bibnamefont {Li}}, \bibinfo {author} {\bibfnamefont {Haijun}\ \bibnamefont
  {Liao}}, \bibinfo {author} {\bibfnamefont {Zhen}\ \bibnamefont {Zhang}},
  \bibinfo {author} {\bibfnamefont {Shiyan}\ \bibnamefont {Li}}, \bibinfo
  {author} {\bibfnamefont {Feng}\ \bibnamefont {Jin}}, \bibinfo {author}
  {\bibfnamefont {Langsheng}\ \bibnamefont {Ling}}, \bibinfo {author}
  {\bibfnamefont {Lei}\ \bibnamefont {Zhang}}, \bibinfo {author} {\bibfnamefont
  {Youming}\ \bibnamefont {Zou}}, \bibinfo {author} {\bibfnamefont
  {Li}~\bibnamefont {Pi}}, \bibinfo {author} {\bibfnamefont {Zhaorong}\
  \bibnamefont {Yang}}, \bibinfo {author} {\bibfnamefont {Junfeng}\
  \bibnamefont {Wang}}, \bibinfo {author} {\bibfnamefont {Zhonghua}\
  \bibnamefont {Wu}}, \ and\ \bibinfo {author} {\bibfnamefont {Qingming}\
  \bibnamefont {Zhang}},\ }\bibfield  {title} {\enquote {\bibinfo {title}
  {{Gapless quantum spin liquid ground state in the two-dimensional spin-1/2
  triangular antiferromagnet YbMgGaO$_4$}},}\ }\href@noop {} {\bibfield
  {journal} {\bibinfo  {journal} {Sci. Rep.}\ }\textbf {\bibinfo {volume}
  {5}},\ \bibinfo {pages} {16419} (\bibinfo {year}
  {2015}{\natexlab{a}})}\BibitemShut {NoStop}%
\bibitem [{\citenamefont {Li}\ \emph {et~al.}(2015{\natexlab{b}})\citenamefont
  {Li}, \citenamefont {Chen}, \citenamefont {Tong}, \citenamefont {Pi},
  \citenamefont {Liu}, \citenamefont {Yang}, \citenamefont {Wang},\ and\
  \citenamefont {Zhang}}]{prl115_167203}%
  \BibitemOpen
  \bibfield  {author} {\bibinfo {author} {\bibfnamefont {Yuesheng}\
  \bibnamefont {Li}}, \bibinfo {author} {\bibfnamefont {Gang}\ \bibnamefont
  {Chen}}, \bibinfo {author} {\bibfnamefont {Wei}\ \bibnamefont {Tong}},
  \bibinfo {author} {\bibfnamefont {Li}~\bibnamefont {Pi}}, \bibinfo {author}
  {\bibfnamefont {Juanjuan}\ \bibnamefont {Liu}}, \bibinfo {author}
  {\bibfnamefont {Zhaorong}\ \bibnamefont {Yang}}, \bibinfo {author}
  {\bibfnamefont {Xiaoqun}\ \bibnamefont {Wang}}, \ and\ \bibinfo {author}
  {\bibfnamefont {Qingming}\ \bibnamefont {Zhang}},\ }\bibfield  {title}
  {\enquote {\bibinfo {title} {{Rare-Earth Triangular Lattice Spin Liquid: A
  Single-Crystal Study of ${\mathrm{YbMgGaO}}_{4}$}},}\ }\href@noop {}
  {\bibfield  {journal} {\bibinfo  {journal} {Phys. Rev. Lett.}\ }\textbf
  {\bibinfo {volume} {115}},\ \bibinfo {pages} {167203} (\bibinfo {year}
  {2015}{\natexlab{b}})}\BibitemShut {NoStop}%
\bibitem [{\citenamefont {Paddison}\ \emph {et~al.}(2017)\citenamefont
  {Paddison}, \citenamefont {Daum}, \citenamefont {Dun}, \citenamefont
  {Ehlers}, \citenamefont {Liu}, \citenamefont {Stone}, \citenamefont {Zhou},\
  and\ \citenamefont {Mourigal}}]{np13_117}%
  \BibitemOpen
  \bibfield  {author} {\bibinfo {author} {\bibfnamefont {Joseph A.~M.}\
  \bibnamefont {Paddison}}, \bibinfo {author} {\bibfnamefont {Marcus}\
  \bibnamefont {Daum}}, \bibinfo {author} {\bibfnamefont {Zhiling}\
  \bibnamefont {Dun}}, \bibinfo {author} {\bibfnamefont {Georg}\ \bibnamefont
  {Ehlers}}, \bibinfo {author} {\bibfnamefont {Yaohua}\ \bibnamefont {Liu}},
  \bibinfo {author} {\bibfnamefont {Matthew~B.}\ \bibnamefont {Stone}},
  \bibinfo {author} {\bibfnamefont {Haidong}\ \bibnamefont {Zhou}}, \ and\
  \bibinfo {author} {\bibfnamefont {Martin}\ \bibnamefont {Mourigal}},\
  }\bibfield  {title} {\enquote {\bibinfo {title} {{Continuous excitations of
  the triangular-lattice quantum spin liquid YbMgGaO$_4$}},}\ }\href@noop {}
  {\bibfield  {journal} {\bibinfo  {journal} {Nat. Phys.}\ }\textbf {\bibinfo
  {volume} {13}},\ \bibinfo {pages} {117} (\bibinfo {year}
  {2017})}\BibitemShut {NoStop}%
\bibitem [{\citenamefont {Shen}\ \emph {et~al.}(2016)\citenamefont {Shen},
  \citenamefont {Li}, \citenamefont {Wo}, \citenamefont {Li}, \citenamefont
  {Shen}, \citenamefont {Pan}, \citenamefont {Wang}, \citenamefont {Walker},
  \citenamefont {Steffens}, \citenamefont {Boehm}, \citenamefont {Hao},
  \citenamefont {Quintero-Castro}, \citenamefont {Harriger}, \citenamefont
  {Frontzek}, \citenamefont {Hao}, \citenamefont {Meng}, \citenamefont {Zhang},
  \citenamefont {Chen},\ and\ \citenamefont {Zhao}}]{nature540_559}%
  \BibitemOpen
  \bibfield  {author} {\bibinfo {author} {\bibfnamefont {Yao}\ \bibnamefont
  {Shen}}, \bibinfo {author} {\bibfnamefont {Yao-Dong}\ \bibnamefont {Li}},
  \bibinfo {author} {\bibfnamefont {Hongliang}\ \bibnamefont {Wo}}, \bibinfo
  {author} {\bibfnamefont {Yuesheng}\ \bibnamefont {Li}}, \bibinfo {author}
  {\bibfnamefont {Shoudong}\ \bibnamefont {Shen}}, \bibinfo {author}
  {\bibfnamefont {Bingying}\ \bibnamefont {Pan}}, \bibinfo {author}
  {\bibfnamefont {Qisi}\ \bibnamefont {Wang}}, \bibinfo {author} {\bibfnamefont
  {H.~C.}\ \bibnamefont {Walker}}, \bibinfo {author} {\bibfnamefont
  {P.}~\bibnamefont {Steffens}}, \bibinfo {author} {\bibfnamefont
  {M.}~\bibnamefont {Boehm}}, \bibinfo {author} {\bibfnamefont {Yiqing}\
  \bibnamefont {Hao}}, \bibinfo {author} {\bibfnamefont {D.~L.}\ \bibnamefont
  {Quintero-Castro}}, \bibinfo {author} {\bibfnamefont {L.~W.}\ \bibnamefont
  {Harriger}}, \bibinfo {author} {\bibfnamefont {M.~D.}\ \bibnamefont
  {Frontzek}}, \bibinfo {author} {\bibfnamefont {Lijie}\ \bibnamefont {Hao}},
  \bibinfo {author} {\bibfnamefont {Siqin}\ \bibnamefont {Meng}}, \bibinfo
  {author} {\bibfnamefont {Qingming}\ \bibnamefont {Zhang}}, \bibinfo {author}
  {\bibfnamefont {Gang}\ \bibnamefont {Chen}}, \ and\ \bibinfo {author}
  {\bibfnamefont {Jun}\ \bibnamefont {Zhao}},\ }\bibfield  {title} {\enquote
  {\bibinfo {title} {{Evidence for a spinon Fermi surface in a
  triangular-lattice quantum-spin-liquid candidate}},}\ }\href@noop {}
  {\bibfield  {journal} {\bibinfo  {journal} {Nature}\ }\textbf {\bibinfo
  {volume} {540}},\ \bibinfo {pages} {559} (\bibinfo {year}
  {2016})}\BibitemShut {NoStop}%
\bibitem [{\citenamefont {Li}\ \emph {et~al.}(2016)\citenamefont {Li},
  \citenamefont {Adroja}, \citenamefont {Biswas}, \citenamefont {Baker},
  \citenamefont {Zhang}, \citenamefont {Liu}, \citenamefont {Tsirlin},
  \citenamefont {Gegenwart},\ and\ \citenamefont
  {Zhang}}]{PhysRevLett.117.097201}%
  \BibitemOpen
  \bibfield  {author} {\bibinfo {author} {\bibfnamefont {Yuesheng}\
  \bibnamefont {Li}}, \bibinfo {author} {\bibfnamefont {Devashibhai}\
  \bibnamefont {Adroja}}, \bibinfo {author} {\bibfnamefont {Pabitra~K.}\
  \bibnamefont {Biswas}}, \bibinfo {author} {\bibfnamefont {Peter~J.}\
  \bibnamefont {Baker}}, \bibinfo {author} {\bibfnamefont {Qian}\ \bibnamefont
  {Zhang}}, \bibinfo {author} {\bibfnamefont {Juanjuan}\ \bibnamefont {Liu}},
  \bibinfo {author} {\bibfnamefont {Alexander~A.}\ \bibnamefont {Tsirlin}},
  \bibinfo {author} {\bibfnamefont {Philipp}\ \bibnamefont {Gegenwart}}, \ and\
  \bibinfo {author} {\bibfnamefont {Qingming}\ \bibnamefont {Zhang}},\
  }\bibfield  {title} {\enquote {\bibinfo {title} {{Muon Spin Relaxation
  Evidence for the U(1) Quantum Spin-Liquid Ground State in the Triangular
  Antiferromagnet YbMgGaO$_{4}$}},}\ }\href {\doibase
  10.1103/PhysRevLett.117.097201} {\bibfield  {journal} {\bibinfo  {journal}
  {Phys. Rev. Lett.}\ }\textbf {\bibinfo {volume} {117}},\ \bibinfo {pages}
  {097201} (\bibinfo {year} {2016})}\BibitemShut {NoStop}%
\bibitem [{\citenamefont {Ma}\ \emph {et~al.}(2018)\citenamefont {Ma},
  \citenamefont {Wang}, \citenamefont {Dong}, \citenamefont {Zhang},
  \citenamefont {Li}, \citenamefont {Zheng}, \citenamefont {Yu}, \citenamefont
  {Wang}, \citenamefont {Che}, \citenamefont {Ran}, \citenamefont {Bao},
  \citenamefont {Cai}, \citenamefont {\ifmmode~\check{C}\else
  \v{C}\fi{}erm\'ak}, \citenamefont {Schneidewind}, \citenamefont {Yano},
  \citenamefont {Gardner}, \citenamefont {Lu}, \citenamefont {Yu},
  \citenamefont {Liu}, \citenamefont {Li}, \citenamefont {Li},\ and\
  \citenamefont {Wen}}]{PhysRevLett.120.087201}%
  \BibitemOpen
  \bibfield  {author} {\bibinfo {author} {\bibfnamefont {Zhen}\ \bibnamefont
  {Ma}}, \bibinfo {author} {\bibfnamefont {Jinghui}\ \bibnamefont {Wang}},
  \bibinfo {author} {\bibfnamefont {Zhao-Yang}\ \bibnamefont {Dong}}, \bibinfo
  {author} {\bibfnamefont {Jun}\ \bibnamefont {Zhang}}, \bibinfo {author}
  {\bibfnamefont {Shichao}\ \bibnamefont {Li}}, \bibinfo {author}
  {\bibfnamefont {Shu-Han}\ \bibnamefont {Zheng}}, \bibinfo {author}
  {\bibfnamefont {Yunjie}\ \bibnamefont {Yu}}, \bibinfo {author} {\bibfnamefont
  {Wei}\ \bibnamefont {Wang}}, \bibinfo {author} {\bibfnamefont {Liqiang}\
  \bibnamefont {Che}}, \bibinfo {author} {\bibfnamefont {Kejing}\ \bibnamefont
  {Ran}}, \bibinfo {author} {\bibfnamefont {Song}\ \bibnamefont {Bao}},
  \bibinfo {author} {\bibfnamefont {Zhengwei}\ \bibnamefont {Cai}}, \bibinfo
  {author} {\bibfnamefont {P.}~\bibnamefont {\ifmmode~\check{C}\else
  \v{C}\fi{}erm\'ak}}, \bibinfo {author} {\bibfnamefont {A.}~\bibnamefont
  {Schneidewind}}, \bibinfo {author} {\bibfnamefont {S.}~\bibnamefont {Yano}},
  \bibinfo {author} {\bibfnamefont {J.~S.}\ \bibnamefont {Gardner}}, \bibinfo
  {author} {\bibfnamefont {Xin}\ \bibnamefont {Lu}}, \bibinfo {author}
  {\bibfnamefont {Shun-Li}\ \bibnamefont {Yu}}, \bibinfo {author}
  {\bibfnamefont {Jun-Ming}\ \bibnamefont {Liu}}, \bibinfo {author}
  {\bibfnamefont {Shiyan}\ \bibnamefont {Li}}, \bibinfo {author} {\bibfnamefont
  {Jian-Xin}\ \bibnamefont {Li}}, \ and\ \bibinfo {author} {\bibfnamefont
  {Jinsheng}\ \bibnamefont {Wen}},\ }\bibfield  {title} {\enquote {\bibinfo
  {title} {{Spin-Glass Ground State in a Triangular-Lattice Compound
  YbZnGaO$_4$}},}\ }\href {\doibase 10.1103/PhysRevLett.120.087201} {\bibfield
  {journal} {\bibinfo  {journal} {Phys. Rev. Lett.}\ }\textbf {\bibinfo
  {volume} {120}},\ \bibinfo {pages} {087201} (\bibinfo {year}
  {2018})}\BibitemShut {NoStop}%
\bibitem [{\citenamefont {Liu}\ \emph {et~al.}(2018{\natexlab{a}})\citenamefont
  {Liu}, \citenamefont {Zhang}, \citenamefont {Ji}, \citenamefont {Liu},
  \citenamefont {Li}, \citenamefont {Wang}, \citenamefont {Lei}, \citenamefont
  {Chen},\ and\ \citenamefont {Zhang}}]{Liu_2018}%
  \BibitemOpen
  \bibfield  {author} {\bibinfo {author} {\bibfnamefont {Weiwei}\ \bibnamefont
  {Liu}}, \bibinfo {author} {\bibfnamefont {Zheng}\ \bibnamefont {Zhang}},
  \bibinfo {author} {\bibfnamefont {Jianting}\ \bibnamefont {Ji}}, \bibinfo
  {author} {\bibfnamefont {Yixuan}\ \bibnamefont {Liu}}, \bibinfo {author}
  {\bibfnamefont {Jianshu}\ \bibnamefont {Li}}, \bibinfo {author}
  {\bibfnamefont {Xiaoqun}\ \bibnamefont {Wang}}, \bibinfo {author}
  {\bibfnamefont {Hechang}\ \bibnamefont {Lei}}, \bibinfo {author}
  {\bibfnamefont {Gang}\ \bibnamefont {Chen}}, \ and\ \bibinfo {author}
  {\bibfnamefont {Qingming}\ \bibnamefont {Zhang}},\ }\bibfield  {title}
  {\enquote {\bibinfo {title} {{Rare-Earth Chalcogenides: A Large Family of
  Triangular Lattice Spin Liquid Candidates}},}\ }\href {\doibase
  10.1088/0256-307x/35/11/117501} {\bibfield  {journal} {\bibinfo  {journal}
  {Chin. Phys. Lett.}\ }\textbf {\bibinfo {volume} {35}},\ \bibinfo
  {pages} {117501} (\bibinfo {year} {2018}{\natexlab{a}})}\BibitemShut
  {NoStop}%
\bibitem [{\citenamefont {Baenitz}\ \emph {et~al.}(2018)\citenamefont
  {Baenitz}, \citenamefont {Schlender}, \citenamefont {Sichelschmidt},
  \citenamefont {Onykiienko}, \citenamefont {Zangeneh}, \citenamefont
  {Ranjith}, \citenamefont {Sarkar}, \citenamefont {Hozoi}, \citenamefont
  {Walker}, \citenamefont {Orain}, \citenamefont {Yasuoka}, \citenamefont
  {van~den Brink}, \citenamefont {Klauss}, \citenamefont {Inosov},\ and\
  \citenamefont {Doert}}]{PhysRevB.98.220409}%
  \BibitemOpen
  \bibfield  {author} {\bibinfo {author} {\bibfnamefont {M.}~\bibnamefont
  {Baenitz}}, \bibinfo {author} {\bibfnamefont {Ph.}\ \bibnamefont
  {Schlender}}, \bibinfo {author} {\bibfnamefont {J.}~\bibnamefont
  {Sichelschmidt}}, \bibinfo {author} {\bibfnamefont {Y.~A.}\ \bibnamefont
  {Onykiienko}}, \bibinfo {author} {\bibfnamefont {Z.}~\bibnamefont
  {Zangeneh}}, \bibinfo {author} {\bibfnamefont {K.~M.}\ \bibnamefont
  {Ranjith}}, \bibinfo {author} {\bibfnamefont {R.}~\bibnamefont {Sarkar}},
  \bibinfo {author} {\bibfnamefont {L.}~\bibnamefont {Hozoi}}, \bibinfo
  {author} {\bibfnamefont {H.~C.}\ \bibnamefont {Walker}}, \bibinfo {author}
  {\bibfnamefont {J.-C.}\ \bibnamefont {Orain}}, \bibinfo {author}
  {\bibfnamefont {H.}~\bibnamefont {Yasuoka}}, \bibinfo {author} {\bibfnamefont
  {J.}~\bibnamefont {van~den Brink}}, \bibinfo {author} {\bibfnamefont {H.~H.}\
  \bibnamefont {Klauss}}, \bibinfo {author} {\bibfnamefont {D.~S.}\
  \bibnamefont {Inosov}}, \ and\ \bibinfo {author} {\bibfnamefont {Th.}\
  \bibnamefont {Doert}},\ }\bibfield  {title} {\enquote {\bibinfo {title}
  {{NaYbS$_2$: A planar spin-1/2 triangular-lattice magnet and putative spin
  liquid}},}\ }\href {\doibase 10.1103/PhysRevB.98.220409} {\bibfield
  {journal} {\bibinfo  {journal} {Phys. Rev. B}\ }\textbf {\bibinfo {volume}
  {98}},\ \bibinfo {pages} {220409} (\bibinfo {year} {2018})}\BibitemShut
  {NoStop}%
\bibitem [{\citenamefont {M.~Bordelon}\ \emph {et~al.}(2019)\citenamefont
  {M.~Bordelon}, \citenamefont {Kenney}, \citenamefont {Liu}, \citenamefont
  {Hogan}, \citenamefont {Posthuma}, \citenamefont {Kavand}, \citenamefont
  {Lyu}, \citenamefont {Sherwin}, \citenamefont {Butch}, \citenamefont {Brown},
  \citenamefont {Graf}, \citenamefont {Balents},\ and\ \citenamefont
  {Wilson}}]{np15_1058}%
  \BibitemOpen
  \bibfield  {author} {\bibinfo {author} {\bibfnamefont {Mitchell}\
  \bibnamefont {M.~Bordelon}}, \bibinfo {author} {\bibfnamefont {Eric}\
  \bibnamefont {Kenney}}, \bibinfo {author} {\bibfnamefont {Chunxiao}\
  \bibnamefont {Liu}}, \bibinfo {author} {\bibfnamefont {Tom}\ \bibnamefont
  {Hogan}}, \bibinfo {author} {\bibfnamefont {Lorenzo}\ \bibnamefont
  {Posthuma}}, \bibinfo {author} {\bibfnamefont {Marzieh}\ \bibnamefont
  {Kavand}}, \bibinfo {author} {\bibfnamefont {Yuanqi}\ \bibnamefont {Lyu}},
  \bibinfo {author} {\bibfnamefont {Mark}\ \bibnamefont {Sherwin}}, \bibinfo
  {author} {\bibfnamefont {N.~P.}\ \bibnamefont {Butch}}, \bibinfo {author}
  {\bibfnamefont {Craig}\ \bibnamefont {Brown}}, \bibinfo {author}
  {\bibfnamefont {M.~J.}\ \bibnamefont {Graf}}, \bibinfo {author}
  {\bibfnamefont {Leon}\ \bibnamefont {Balents}}, \ and\ \bibinfo {author}
  {\bibfnamefont {Stephen~D.}\ \bibnamefont {Wilson}},\ }\bibfield  {title}
  {\enquote {\bibinfo {title} {{Field-tunable quantum disordered ground state
  in the triangular lattice antiferromagnet NaYbO$_2$}},}\ }\href@noop {}
  {\bibfield  {journal} {\bibinfo  {journal} {Nat. Phys.}\ }\textbf
  {\bibinfo {volume} {15}},\ \bibinfo {pages} {1058} (\bibinfo {year}
  {2019})}\BibitemShut {NoStop}%
\bibitem [{\citenamefont {Ding}\ \emph {et~al.}(2019)\citenamefont {Ding},
  \citenamefont {Manuel}, \citenamefont {Bachus}, \citenamefont {Gru\ss{}ler},
  \citenamefont {Gegenwart}, \citenamefont {Singleton}, \citenamefont
  {Johnson}, \citenamefont {Walker}, \citenamefont {Adroja}, \citenamefont
  {Hillier},\ and\ \citenamefont {Tsirlin}}]{PhysRevB.100.144432}%
  \BibitemOpen
  \bibfield  {author} {\bibinfo {author} {\bibfnamefont {Lei}\ \bibnamefont
  {Ding}}, \bibinfo {author} {\bibfnamefont {Pascal}\ \bibnamefont {Manuel}},
  \bibinfo {author} {\bibfnamefont {Sebastian}\ \bibnamefont {Bachus}},
  \bibinfo {author} {\bibfnamefont {Franziska}\ \bibnamefont {Gru\ss{}ler}},
  \bibinfo {author} {\bibfnamefont {Philipp}\ \bibnamefont {Gegenwart}},
  \bibinfo {author} {\bibfnamefont {John}\ \bibnamefont {Singleton}}, \bibinfo
  {author} {\bibfnamefont {Roger~D.}\ \bibnamefont {Johnson}}, \bibinfo
  {author} {\bibfnamefont {Helen~C.}\ \bibnamefont {Walker}}, \bibinfo {author}
  {\bibfnamefont {Devashibhai~T.}\ \bibnamefont {Adroja}}, \bibinfo {author}
  {\bibfnamefont {Adrian~D.}\ \bibnamefont {Hillier}}, \ and\ \bibinfo {author}
  {\bibfnamefont {Alexander~A.}\ \bibnamefont {Tsirlin}},\ }\bibfield  {title}
  {\enquote {\bibinfo {title} {{Gapless spin-liquid state in the structurally
  disorder-free triangular antiferromagnet NaYbO$_2$}},}\ }\href {\doibase
  10.1103/PhysRevB.100.144432} {\bibfield  {journal} {\bibinfo  {journal}
  {Phys. Rev. B}\ }\textbf {\bibinfo {volume} {100}},\ \bibinfo {pages}
  {144432} (\bibinfo {year} {2019})}\BibitemShut {NoStop}%
\bibitem [{\citenamefont {Sarkar}\ \emph {et~al.}(2019)\citenamefont {Sarkar},
  \citenamefont {Schlender}, \citenamefont {Grinenko}, \citenamefont
  {Haeussler}, \citenamefont {Baker}, \citenamefont {Doert},\ and\
  \citenamefont {Klauss}}]{PhysRevB.100.241116}%
  \BibitemOpen
  \bibfield  {author} {\bibinfo {author} {\bibfnamefont {R.}~\bibnamefont
  {Sarkar}}, \bibinfo {author} {\bibfnamefont {Ph.}\ \bibnamefont {Schlender}},
  \bibinfo {author} {\bibfnamefont {V.}~\bibnamefont {Grinenko}}, \bibinfo
  {author} {\bibfnamefont {E.}~\bibnamefont {Haeussler}}, \bibinfo {author}
  {\bibfnamefont {Peter~J.}\ \bibnamefont {Baker}}, \bibinfo {author}
  {\bibfnamefont {Th.}\ \bibnamefont {Doert}}, \ and\ \bibinfo {author}
  {\bibfnamefont {H.-H.}\ \bibnamefont {Klauss}},\ }\bibfield  {title}
  {\enquote {\bibinfo {title} {{Quantum spin liquid ground state in the
  disorder free triangular lattice NaYbS$_2$}},}\ }\href {\doibase
  10.1103/PhysRevB.100.241116} {\bibfield  {journal} {\bibinfo  {journal}
  {Phys. Rev. B}\ }\textbf {\bibinfo {volume} {100}},\ \bibinfo {pages}
  {241116} (\bibinfo {year} {2019})}\BibitemShut {NoStop}%
\bibitem [{\citenamefont {Ranjith}\ \emph {et~al.}(2019)\citenamefont
  {Ranjith}, \citenamefont {Luther}, \citenamefont {Reimann}, \citenamefont
  {Schmidt}, \citenamefont {Schlender}, \citenamefont {Sichelschmidt},
  \citenamefont {Yasuoka}, \citenamefont {Strydom}, \citenamefont {Skourski},
  \citenamefont {Wosnitza}, \citenamefont {K\"uhne}, \citenamefont {Doert},\
  and\ \citenamefont {Baenitz}}]{PhysRevB.100.224417}%
  \BibitemOpen
  \bibfield  {author} {\bibinfo {author} {\bibfnamefont {K.~M.}\ \bibnamefont
  {Ranjith}}, \bibinfo {author} {\bibfnamefont {S.}~\bibnamefont {Luther}},
  \bibinfo {author} {\bibfnamefont {T.}~\bibnamefont {Reimann}}, \bibinfo
  {author} {\bibfnamefont {B.}~\bibnamefont {Schmidt}}, \bibinfo {author}
  {\bibfnamefont {Ph.}\ \bibnamefont {Schlender}}, \bibinfo {author}
  {\bibfnamefont {J.}~\bibnamefont {Sichelschmidt}}, \bibinfo {author}
  {\bibfnamefont {H.}~\bibnamefont {Yasuoka}}, \bibinfo {author} {\bibfnamefont
  {A.~M.}\ \bibnamefont {Strydom}}, \bibinfo {author} {\bibfnamefont
  {Y.}~\bibnamefont {Skourski}}, \bibinfo {author} {\bibfnamefont
  {J.}~\bibnamefont {Wosnitza}}, \bibinfo {author} {\bibfnamefont
  {H.}~\bibnamefont {K\"uhne}}, \bibinfo {author} {\bibfnamefont {Th.}\
  \bibnamefont {Doert}}, \ and\ \bibinfo {author} {\bibfnamefont
  {M.}~\bibnamefont {Baenitz}},\ }\bibfield  {title} {\enquote {\bibinfo
  {title} {{Anisotropic field-induced ordering in the triangular-lattice
  quantum spin liquid NaYbSe$_2$}},}\ }\href {\doibase
  10.1103/PhysRevB.100.224417} {\bibfield  {journal} {\bibinfo  {journal}
  {Phys. Rev. B}\ }\textbf {\bibinfo {volume} {100}},\ \bibinfo {pages}
  {224417} (\bibinfo {year} {2019})}\BibitemShut {NoStop}%
\bibitem [{\citenamefont {Xing}\ \emph {et~al.}(2020)\citenamefont {Xing},
  \citenamefont {Sanjeewa}, \citenamefont {Kim}, \citenamefont {Stewart},
  \citenamefont {Du}, \citenamefont {Reboredo}, \citenamefont {Custelcean},\
  and\ \citenamefont {Sefat}}]{doi:10.1021/acsmaterialslett.9b00464}%
  \BibitemOpen
  \bibfield  {author} {\bibinfo {author} {\bibfnamefont {Jie}\ \bibnamefont
  {Xing}}, \bibinfo {author} {\bibfnamefont {Liurukara~D.}\ \bibnamefont
  {Sanjeewa}}, \bibinfo {author} {\bibfnamefont {Jungsoo}\ \bibnamefont {Kim}},
  \bibinfo {author} {\bibfnamefont {G.~R.}\ \bibnamefont {Stewart}}, \bibinfo
  {author} {\bibfnamefont {Mao-Hua}\ \bibnamefont {Du}}, \bibinfo {author}
  {\bibfnamefont {Fernando~A.}\ \bibnamefont {Reboredo}}, \bibinfo {author}
  {\bibfnamefont {Radu}\ \bibnamefont {Custelcean}}, \ and\ \bibinfo {author}
  {\bibfnamefont {Athena~S.}\ \bibnamefont {Sefat}},\ }\bibfield  {title}
  {\enquote {\bibinfo {title} {{Crystal Synthesis and Frustrated Magnetism in
  Triangular Lattice Cs$RE$Se$_2$ ($RE$ = La-Lu): Quantum Spin Liquid
  Candidates CsCeSe$_2$ and CsYbSe$_2$}},}\ }\href {\doibase
  10.1021/acsmaterialslett.9b00464} {\bibfield  {journal} {\bibinfo  {journal}
  {ACS Mater. Lett.}\ }\textbf {\bibinfo {volume} {2}},\ \bibinfo {pages}
  {71} (\bibinfo {year} {2020})}\BibitemShut {NoStop}%
\bibitem [{\citenamefont {Xing}\ \emph
  {et~al.}(2019{\natexlab{a}})\citenamefont {Xing}, \citenamefont {Sanjeewa},
  \citenamefont {Kim}, \citenamefont {Meier}, \citenamefont {May},
  \citenamefont {Zheng}, \citenamefont {Custelcean}, \citenamefont {Stewart},\
  and\ \citenamefont {Sefat}}]{PhysRevMaterials.3.114413}%
  \BibitemOpen
  \bibfield  {author} {\bibinfo {author} {\bibfnamefont {Jie}\ \bibnamefont
  {Xing}}, \bibinfo {author} {\bibfnamefont {Liurukara~D.}\ \bibnamefont
  {Sanjeewa}}, \bibinfo {author} {\bibfnamefont {Jungsoo}\ \bibnamefont {Kim}},
  \bibinfo {author} {\bibfnamefont {William~R.}\ \bibnamefont {Meier}},
  \bibinfo {author} {\bibfnamefont {Andrew~F.}\ \bibnamefont {May}}, \bibinfo
  {author} {\bibfnamefont {Qiang}\ \bibnamefont {Zheng}}, \bibinfo {author}
  {\bibfnamefont {Radu}\ \bibnamefont {Custelcean}}, \bibinfo {author}
  {\bibfnamefont {G.~R.}\ \bibnamefont {Stewart}}, \ and\ \bibinfo {author}
  {\bibfnamefont {Athena~S.}\ \bibnamefont {Sefat}},\ }\bibfield  {title}
  {\enquote {\bibinfo {title} {{Synthesis, magnetization, and heat capacity of
  triangular lattice materials NaErSe$_2$ and KErSe$_2$}},}\ }\href {\doibase
  10.1103/PhysRevMaterials.3.114413} {\bibfield  {journal} {\bibinfo  {journal}
  {Phys. Rev. Mater.}\ }\textbf {\bibinfo {volume} {3}},\ \bibinfo {pages}
  {114413} (\bibinfo {year} {2019}{\natexlab{a}})}\BibitemShut {NoStop}%
\bibitem [{\citenamefont {Xing}\ \emph
  {et~al.}(2019{\natexlab{b}})\citenamefont {Xing}, \citenamefont {Sanjeewa},
  \citenamefont {Kim}, \citenamefont {Stewart}, \citenamefont {Podlesnyak},\
  and\ \citenamefont {Sefat}}]{PhysRevB.100.220407}%
  \BibitemOpen
  \bibfield  {author} {\bibinfo {author} {\bibfnamefont {Jie}\ \bibnamefont
  {Xing}}, \bibinfo {author} {\bibfnamefont {Liurukara~D.}\ \bibnamefont
  {Sanjeewa}}, \bibinfo {author} {\bibfnamefont {Jungsoo}\ \bibnamefont {Kim}},
  \bibinfo {author} {\bibfnamefont {G.~R.}\ \bibnamefont {Stewart}}, \bibinfo
  {author} {\bibfnamefont {Andrey}\ \bibnamefont {Podlesnyak}}, \ and\ \bibinfo
  {author} {\bibfnamefont {Athena~S.}\ \bibnamefont {Sefat}},\ }\bibfield
  {title} {\enquote {\bibinfo {title} {{Field-induced magnetic transition and
  spin fluctuations in the quantum spin-liquid candidate CsYbSe$_2$}},}\ }\href
  {\doibase 10.1103/PhysRevB.100.220407} {\bibfield  {journal} {\bibinfo
  {journal} {Phys. Rev. B}\ }\textbf {\bibinfo {volume} {100}},\ \bibinfo
  {pages} {220407} (\bibinfo {year} {2019}{\natexlab{b}})}\BibitemShut
  {NoStop}%
\bibitem [{\citenamefont {Han}\ \emph {et~al.}(2016)\citenamefont {Han},
  \citenamefont {Norman}, \citenamefont {Wen}, \citenamefont
  {Rodriguez-Rivera}, \citenamefont {Helton}, \citenamefont {Broholm},\ and\
  \citenamefont {Lee}}]{PhysRevB.94.060409}%
  \BibitemOpen
  \bibfield  {author} {\bibinfo {author} {\bibfnamefont {Tian-Heng}\
  \bibnamefont {Han}}, \bibinfo {author} {\bibfnamefont {M.~R.}\ \bibnamefont
  {Norman}}, \bibinfo {author} {\bibfnamefont {J.-J.}\ \bibnamefont {Wen}},
  \bibinfo {author} {\bibfnamefont {Jose~A.}\ \bibnamefont {Rodriguez-Rivera}},
  \bibinfo {author} {\bibfnamefont {Joel~S.}\ \bibnamefont {Helton}}, \bibinfo
  {author} {\bibfnamefont {Collin}\ \bibnamefont {Broholm}}, \ and\ \bibinfo
  {author} {\bibfnamefont {Young~S.}\ \bibnamefont {Lee}},\ }\bibfield  {title}
  {\enquote {\bibinfo {title} {Correlated impurities and intrinsic spin-liquid
  physics in the kagome material herbertsmithite},}\ }\href {\doibase
  10.1103/PhysRevB.94.060409} {\bibfield  {journal} {\bibinfo  {journal} {Phys.
  Rev. B}\ }\textbf {\bibinfo {volume} {94}},\ \bibinfo {pages} {060409}
  (\bibinfo {year} {2016})}\BibitemShut {NoStop}%
\bibitem [{\citenamefont {Zhu}\ \emph {et~al.}(2017)\citenamefont {Zhu},
  \citenamefont {Maksimov}, \citenamefont {White},\ and\ \citenamefont
  {Chernyshev}}]{PhysRevLett.119.157201}%
  \BibitemOpen
  \bibfield  {author} {\bibinfo {author} {\bibfnamefont {Zhenyue}\ \bibnamefont
  {Zhu}}, \bibinfo {author} {\bibfnamefont {P.~A.}\ \bibnamefont {Maksimov}},
  \bibinfo {author} {\bibfnamefont {Steven~R.}\ \bibnamefont {White}}, \ and\
  \bibinfo {author} {\bibfnamefont {A.~L.}\ \bibnamefont {Chernyshev}},\
  }\bibfield  {title} {\enquote {\bibinfo {title} {{Disorder-Induced Mimicry of
  a Spin Liquid in YbMgGaO$_4$}},}\ }\href {\doibase
  10.1103/PhysRevLett.119.157201} {\bibfield  {journal} {\bibinfo  {journal}
  {Phys. Rev. Lett.}\ }\textbf {\bibinfo {volume} {119}},\ \bibinfo {pages}
  {157201} (\bibinfo {year} {2017})}\BibitemShut {NoStop}%
\bibitem [{\citenamefont {Parker}\ and\ \citenamefont
  {Balents}(2018)}]{PhysRevB.97.184413}%
  \BibitemOpen
  \bibfield  {author} {\bibinfo {author} {\bibfnamefont {Edward}\ \bibnamefont
  {Parker}}\ and\ \bibinfo {author} {\bibfnamefont {Leon}\ \bibnamefont
  {Balents}},\ }\bibfield  {title} {\enquote {\bibinfo {title}
  {{Finite-temperature behavior of a classical spin-orbit-coupled model for
  ${\mathrm{YbMgGaO}}_{4}$ with and without bond disorder}},}\ }\href {\doibase
  10.1103/PhysRevB.97.184413} {\bibfield  {journal} {\bibinfo  {journal} {Phys.
  Rev. B}\ }\textbf {\bibinfo {volume} {97}},\ \bibinfo {pages} {184413}
  (\bibinfo {year} {2018})}\BibitemShut {NoStop}%
\bibitem [{\citenamefont {Li}\ \emph {et~al.}(2017)\citenamefont {Li},
  \citenamefont {Adroja}, \citenamefont {Bewley}, \citenamefont {Voneshen},
  \citenamefont {Tsirlin}, \citenamefont {Gegenwart},\ and\ \citenamefont
  {Zhang}}]{PhysRevLett.118.107202}%
  \BibitemOpen
  \bibfield  {author} {\bibinfo {author} {\bibfnamefont {Yuesheng}\
  \bibnamefont {Li}}, \bibinfo {author} {\bibfnamefont {Devashibhai}\
  \bibnamefont {Adroja}}, \bibinfo {author} {\bibfnamefont {Robert~I.}\
  \bibnamefont {Bewley}}, \bibinfo {author} {\bibfnamefont {David}\
  \bibnamefont {Voneshen}}, \bibinfo {author} {\bibfnamefont {Alexander~A.}\
  \bibnamefont {Tsirlin}}, \bibinfo {author} {\bibfnamefont {Philipp}\
  \bibnamefont {Gegenwart}}, \ and\ \bibinfo {author} {\bibfnamefont
  {Qingming}\ \bibnamefont {Zhang}},\ }\bibfield  {title} {\enquote {\bibinfo
  {title} {{Crystalline Electric-Field Randomness in the Triangular Lattice
  Spin-Liquid YbMgGaO$_4$}},}\ }\href {\doibase 10.1103/PhysRevLett.118.107202}
  {\bibfield  {journal} {\bibinfo  {journal} {Phys. Rev. Lett.}\ }\textbf
  {\bibinfo {volume} {118}},\ \bibinfo {pages} {107202} (\bibinfo {year}
  {2017})}\BibitemShut {NoStop}%
\bibitem [{\citenamefont {Savary}\ and\ \citenamefont
  {Balents}(2017{\natexlab{b}})}]{PhysRevLett.118.087203}%
  \BibitemOpen
  \bibfield  {author} {\bibinfo {author} {\bibfnamefont {Lucile}\ \bibnamefont
  {Savary}}\ and\ \bibinfo {author} {\bibfnamefont {Leon}\ \bibnamefont
  {Balents}},\ }\bibfield  {title} {\enquote {\bibinfo {title}
  {{Disorder-Induced Quantum Spin Liquid in Spin Ice Pyrochlores}},}\ }\href
  {\doibase 10.1103/PhysRevLett.118.087203} {\bibfield  {journal} {\bibinfo
  {journal} {Phys. Rev. Lett.}\ }\textbf {\bibinfo {volume} {118}},\ \bibinfo
  {pages} {087203} (\bibinfo {year} {2017}{\natexlab{b}})}\BibitemShut
  {NoStop}%
\bibitem [{\citenamefont {Zhu}\ \emph {et~al.}(2018)\citenamefont {Zhu},
  \citenamefont {Maksimov}, \citenamefont {White},\ and\ \citenamefont
  {Chernyshev}}]{PhysRevLett.120.207203}%
  \BibitemOpen
  \bibfield  {author} {\bibinfo {author} {\bibfnamefont {Zhenyue}\ \bibnamefont
  {Zhu}}, \bibinfo {author} {\bibfnamefont {P.~A.}\ \bibnamefont {Maksimov}},
  \bibinfo {author} {\bibfnamefont {Steven~R.}\ \bibnamefont {White}}, \ and\
  \bibinfo {author} {\bibfnamefont {A.~L.}\ \bibnamefont {Chernyshev}},\
  }\bibfield  {title} {\enquote {\bibinfo {title} {{Topography of Spin Liquids
  on a Triangular Lattice}},}\ }\href {\doibase 10.1103/PhysRevLett.120.207203}
  {\bibfield  {journal} {\bibinfo  {journal} {Phys. Rev. Lett.}\ }\textbf
  {\bibinfo {volume} {120}},\ \bibinfo {pages} {207203} (\bibinfo {year}
  {2018})}\BibitemShut {NoStop}%
\bibitem [{\citenamefont {Kimchi}\ \emph
  {et~al.}(2018{\natexlab{a}})\citenamefont {Kimchi}, \citenamefont
  {Sheckelton}, \citenamefont {McQueen},\ and\ \citenamefont
  {Lee}}]{nc_Itamar}%
  \BibitemOpen
  \bibfield  {author} {\bibinfo {author} {\bibfnamefont {Itamar}\ \bibnamefont
  {Kimchi}}, \bibinfo {author} {\bibfnamefont {John~P.}\ \bibnamefont
  {Sheckelton}}, \bibinfo {author} {\bibfnamefont {Tyrel~M.}\ \bibnamefont
  {McQueen}}, \ and\ \bibinfo {author} {\bibfnamefont {Patrick~A.}\
  \bibnamefont {Lee}},\ }\bibfield  {title} {\enquote {\bibinfo {title}
  {Scaling and data collapse from local moments in frustrated disordered
  quantum spin systems},}\ }\href@noop {} {\bibfield  {journal} {\bibinfo
  {journal} {Nat. Commun.}\ }\textbf {\bibinfo {volume} {9}},\ \bibinfo
  {pages} {4367} (\bibinfo {year} {2018}{\natexlab{a}})}\BibitemShut {NoStop}%
\bibitem [{\citenamefont {Zhang}\ \emph {et~al.}(2018)\citenamefont {Zhang},
  \citenamefont {Mahmood}, \citenamefont {Daum}, \citenamefont {Dun},
  \citenamefont {Paddison}, \citenamefont {Laurita}, \citenamefont {Hong},
  \citenamefont {Zhou}, \citenamefont {Armitage},\ and\ \citenamefont
  {Mourigal}}]{PhysRevX.8.031001}%
  \BibitemOpen
  \bibfield  {author} {\bibinfo {author} {\bibfnamefont {Xinshu}\ \bibnamefont
  {Zhang}}, \bibinfo {author} {\bibfnamefont {Fahad}\ \bibnamefont {Mahmood}},
  \bibinfo {author} {\bibfnamefont {Marcus}\ \bibnamefont {Daum}}, \bibinfo
  {author} {\bibfnamefont {Zhiling}\ \bibnamefont {Dun}}, \bibinfo {author}
  {\bibfnamefont {Joseph A.~M.}\ \bibnamefont {Paddison}}, \bibinfo {author}
  {\bibfnamefont {Nicholas~J.}\ \bibnamefont {Laurita}}, \bibinfo {author}
  {\bibfnamefont {Tao}\ \bibnamefont {Hong}}, \bibinfo {author} {\bibfnamefont
  {Haidong}\ \bibnamefont {Zhou}}, \bibinfo {author} {\bibfnamefont {N.~P.}\
  \bibnamefont {Armitage}}, \ and\ \bibinfo {author} {\bibfnamefont {Martin}\
  \bibnamefont {Mourigal}},\ }\bibfield  {title} {\enquote {\bibinfo {title}
  {{Hierarchy of Exchange Interactions in the Triangular-Lattice Spin Liquid
  YbMgGaO$_4$}},}\ }\href {\doibase 10.1103/PhysRevX.8.031001} {\bibfield
  {journal} {\bibinfo  {journal} {Phys. Rev. X}\ }\textbf {\bibinfo {volume}
  {8}},\ \bibinfo {pages} {031001} (\bibinfo {year} {2018})}\BibitemShut
  {NoStop}%
\bibitem [{\citenamefont {Kimchi}\ \emph
  {et~al.}(2018{\natexlab{b}})\citenamefont {Kimchi}, \citenamefont {Nahum},\
  and\ \citenamefont {Senthil}}]{PhysRevX.8.031028}%
  \BibitemOpen
  \bibfield  {author} {\bibinfo {author} {\bibfnamefont {Itamar}\ \bibnamefont
  {Kimchi}}, \bibinfo {author} {\bibfnamefont {Adam}\ \bibnamefont {Nahum}}, \
  and\ \bibinfo {author} {\bibfnamefont {T.}~\bibnamefont {Senthil}},\
  }\bibfield  {title} {\enquote {\bibinfo {title} {{Valence Bonds in Random
  Quantum Magnets: Theory and Application to YbMgGaO$_4$}},}\ }\href {\doibase
  10.1103/PhysRevX.8.031028} {\bibfield  {journal} {\bibinfo  {journal} {Phys.
  Rev. X}\ }\textbf {\bibinfo {volume} {8}},\ \bibinfo {pages} {031028}
  (\bibinfo {year} {2018}{\natexlab{b}})}\BibitemShut {NoStop}%
\bibitem [{\citenamefont {Liu}\ \emph {et~al.}(2018{\natexlab{b}})\citenamefont
  {Liu}, \citenamefont {Shao}, \citenamefont {Lin}, \citenamefont {Guo},\ and\
  \citenamefont {Sandvik}}]{PhysRevX.8.041040}%
  \BibitemOpen
  \bibfield  {author} {\bibinfo {author} {\bibfnamefont {Lu}~\bibnamefont
  {Liu}}, \bibinfo {author} {\bibfnamefont {Hui}\ \bibnamefont {Shao}},
  \bibinfo {author} {\bibfnamefont {Yu-Cheng}\ \bibnamefont {Lin}}, \bibinfo
  {author} {\bibfnamefont {Wenan}\ \bibnamefont {Guo}}, \ and\ \bibinfo
  {author} {\bibfnamefont {Anders~W.}\ \bibnamefont {Sandvik}},\ }\bibfield
  {title} {\enquote {\bibinfo {title} {{Random-Singlet Phase in Disordered
  Two-Dimensional Quantum Magnets}},}\ }\href {\doibase
  10.1103/PhysRevX.8.041040} {\bibfield  {journal} {\bibinfo  {journal} {Phys.
  Rev. X}\ }\textbf {\bibinfo {volume} {8}},\ \bibinfo {pages} {041040}
  (\bibinfo {year} {2018}{\natexlab{b}})}\BibitemShut {NoStop}%
\bibitem [{\citenamefont {Uematsu}\ and\ \citenamefont
  {Kawamura}(2019)}]{PhysRevLett.123.087201}%
  \BibitemOpen
  \bibfield  {author} {\bibinfo {author} {\bibfnamefont {Kazuki}\ \bibnamefont
  {Uematsu}}\ and\ \bibinfo {author} {\bibfnamefont {Hikaru}\ \bibnamefont
  {Kawamura}},\ }\bibfield  {title} {\enquote {\bibinfo {title}
  {{Randomness-Induced Quantum Spin Liquid Behavior in the $s=1/2$ Random-Bond
  Heisenberg Antiferromagnet on the Pyrochlore Lattice}},}\ }\href {\doibase
  10.1103/PhysRevLett.123.087201} {\bibfield  {journal} {\bibinfo  {journal}
  {Phys. Rev. Lett.}\ }\textbf {\bibinfo {volume} {123}},\ \bibinfo {pages}
  {087201} (\bibinfo {year} {2019})}\BibitemShut {NoStop}%
\bibitem [{\citenamefont {Shamblin}\ \emph {et~al.}(2017)\citenamefont
  {Shamblin}, \citenamefont {Dun}, \citenamefont {Lee}, \citenamefont
  {Johnston}, \citenamefont {Choi}, \citenamefont {Page}, \citenamefont {Qiu},\
  and\ \citenamefont {Zhou}}]{PhysRevB.96.174418}%
  \BibitemOpen
  \bibfield  {author} {\bibinfo {author} {\bibfnamefont {Jacob}\ \bibnamefont
  {Shamblin}}, \bibinfo {author} {\bibfnamefont {Zhiling}\ \bibnamefont {Dun}},
  \bibinfo {author} {\bibfnamefont {Minseong}\ \bibnamefont {Lee}}, \bibinfo
  {author} {\bibfnamefont {Steve}\ \bibnamefont {Johnston}}, \bibinfo {author}
  {\bibfnamefont {Eun~Sang}\ \bibnamefont {Choi}}, \bibinfo {author}
  {\bibfnamefont {Katharine}\ \bibnamefont {Page}}, \bibinfo {author}
  {\bibfnamefont {Yiming}\ \bibnamefont {Qiu}}, \ and\ \bibinfo {author}
  {\bibfnamefont {Haidong}\ \bibnamefont {Zhou}},\ }\bibfield  {title}
  {\enquote {\bibinfo {title} {{Structural and magnetic short-range order in
  fluorite Yb$_2$TiO$_5$}},}\ }\href {\doibase 10.1103/PhysRevB.96.174418}
  {\bibfield  {journal} {\bibinfo  {journal} {Phys. Rev. B}\ }\textbf {\bibinfo
  {volume} {96}},\ \bibinfo {pages} {174418} (\bibinfo {year}
  {2017})}\BibitemShut {NoStop}%
\bibitem [{\citenamefont {Furukawa}\ \emph {et~al.}(2015)\citenamefont
  {Furukawa}, \citenamefont {Miyagawa}, \citenamefont {Itou}, \citenamefont
  {Ito}, \citenamefont {Taniguchi}, \citenamefont {Saito}, \citenamefont
  {Iguchi}, \citenamefont {Sasaki},\ and\ \citenamefont
  {Kanoda}}]{PhysRevLett.115.077001}%
  \BibitemOpen
  \bibfield  {author} {\bibinfo {author} {\bibfnamefont {T.}~\bibnamefont
  {Furukawa}}, \bibinfo {author} {\bibfnamefont {K.}~\bibnamefont {Miyagawa}},
  \bibinfo {author} {\bibfnamefont {T.}~\bibnamefont {Itou}}, \bibinfo {author}
  {\bibfnamefont {M.}~\bibnamefont {Ito}}, \bibinfo {author} {\bibfnamefont
  {H.}~\bibnamefont {Taniguchi}}, \bibinfo {author} {\bibfnamefont
  {M.}~\bibnamefont {Saito}}, \bibinfo {author} {\bibfnamefont
  {S.}~\bibnamefont {Iguchi}}, \bibinfo {author} {\bibfnamefont
  {T.}~\bibnamefont {Sasaki}}, \ and\ \bibinfo {author} {\bibfnamefont
  {K.}~\bibnamefont {Kanoda}},\ }\bibfield  {title} {\enquote {\bibinfo {title}
  {Quantum Spin Liquid Emerging from Antiferromagnetic Order by Introducing
  Disorder},}\ }\href {\doibase 10.1103/PhysRevLett.115.077001} {\bibfield
  {journal} {\bibinfo  {journal} {Phys. Rev. Lett.}\ }\textbf {\bibinfo
  {volume} {115}},\ \bibinfo {pages} {077001} (\bibinfo {year}
  {2015})}\BibitemShut {NoStop}%
\bibitem [{\citenamefont {Sanders}\ \emph
  {et~al.}(2016{\natexlab{a}})\citenamefont {Sanders}, \citenamefont {Baroudi},
  \citenamefont {Krizan}, \citenamefont {Mukadam},\ and\ \citenamefont
  {Cava}}]{doi:10.1002/pssb.201600256}%
  \BibitemOpen
  \bibfield  {author} {\bibinfo {author} {\bibfnamefont {M.~B.}\ \bibnamefont
  {Sanders}}, \bibinfo {author} {\bibfnamefont {K.~M.}\ \bibnamefont
  {Baroudi}}, \bibinfo {author} {\bibfnamefont {J.~W.}\ \bibnamefont {Krizan}},
  \bibinfo {author} {\bibfnamefont {O.~A.}\ \bibnamefont {Mukadam}}, \ and\
  \bibinfo {author} {\bibfnamefont {R.~J.}\ \bibnamefont {Cava}},\ }\bibfield
  {title} {\enquote {\bibinfo {title} {{Synthesis, crystal structure, and
  magnetic properties of novel 2D kagome materials RE$_3$Sb$_3$Mg$_2$O$_{14}$
  (RE = La, Pr, Sm, Eu, Tb, Ho): Comparison to RE$_3$Sb$_3$Zn$_2$O$_{14}$
  family}},}\ }\href {\doibase 10.1002/pssb.201600256} {\bibfield  {journal}
  {\bibinfo  {journal} {Phys. Status Solidi B}\ }\textbf {\bibinfo {volume}
  {253}},\ \bibinfo {pages} {2056} (\bibinfo {year}
  {2016}{\natexlab{a}})}\BibitemShut {NoStop}%
\bibitem [{\citenamefont {Sanders}\ \emph
  {et~al.}(2016{\natexlab{b}})\citenamefont {Sanders}, \citenamefont {Krizan},\
  and\ \citenamefont {Cava}}]{Sanders2016RE}%
  \BibitemOpen
  \bibfield  {author} {\bibinfo {author} {\bibfnamefont {Marisa~B.}\
  \bibnamefont {Sanders}}, \bibinfo {author} {\bibfnamefont {Jason~W.}\
  \bibnamefont {Krizan}}, \ and\ \bibinfo {author} {\bibfnamefont {Robert~J.}\
  \bibnamefont {Cava}},\ }\bibfield  {title} {\enquote {\bibinfo {title}
  {{RE$_3$Sb$_3$Zn$_2$O$_{14}$ (RE=La, Pr, Nd, Sm, Eu, Gd): A new family of
  pyrochlore derivatives with rare earth ions on a 2D Kagome lattice}},}\
  }\href@noop {} {\bibfield  {journal} {\bibinfo  {journal} {J.
  Mater. Chem. C}\ }\textbf {\bibinfo {volume} {4}},\ \bibinfo {pages}
  {541} (\bibinfo {year} {2016}{\natexlab{b}})}\BibitemShut {NoStop}%
\bibitem [{\citenamefont {Dun}\ \emph {et~al.}(2017)\citenamefont {Dun},
  \citenamefont {Trinh}, \citenamefont {Lee}, \citenamefont {Choi},
  \citenamefont {Li}, \citenamefont {Hu}, \citenamefont {Wang}, \citenamefont
  {Blanc}, \citenamefont {Ramirez},\ and\ \citenamefont
  {Zhou}}]{PhysRevB.95.104439}%
  \BibitemOpen
  \bibfield  {author} {\bibinfo {author} {\bibfnamefont {Z.~L.}\ \bibnamefont
  {Dun}}, \bibinfo {author} {\bibfnamefont {J.}~\bibnamefont {Trinh}}, \bibinfo
  {author} {\bibfnamefont {M.}~\bibnamefont {Lee}}, \bibinfo {author}
  {\bibfnamefont {E.~S.}\ \bibnamefont {Choi}}, \bibinfo {author}
  {\bibfnamefont {K.}~\bibnamefont {Li}}, \bibinfo {author} {\bibfnamefont
  {Y.~F.}\ \bibnamefont {Hu}}, \bibinfo {author} {\bibfnamefont {Y.~X.}\
  \bibnamefont {Wang}}, \bibinfo {author} {\bibfnamefont {N.}~\bibnamefont
  {Blanc}}, \bibinfo {author} {\bibfnamefont {A.~P.}\ \bibnamefont {Ramirez}},
  \ and\ \bibinfo {author} {\bibfnamefont {H.~D.}\ \bibnamefont {Zhou}},\
  }\bibfield  {title} {\enquote {\bibinfo {title} {{Structural and magnetic
  properties of two branches of the tripod-kagome-lattice family
  $A_2R_3$Sb$_3$O$_{14}$ ($A$ = Mg, Zn; $R$ = Pr, Nd, Gd, Tb, Dy, Ho, Er,
  Yb)}},}\ }\href {\doibase 10.1103/PhysRevB.95.104439} {\bibfield  {journal}
  {\bibinfo  {journal} {Phys. Rev. B}\ }\textbf {\bibinfo {volume} {95}},\
  \bibinfo {pages} {104439} (\bibinfo {year} {2017})}\BibitemShut {NoStop}%
\bibitem [{\citenamefont {Ding}\ \emph {et~al.}(2018)\citenamefont {Ding},
  \citenamefont {Yang}, \citenamefont {Zhang}, \citenamefont {Tan},
  \citenamefont {Zhu}, \citenamefont {Chen},\ and\ \citenamefont
  {Shu}}]{PhysRevB.98.174404}%
  \BibitemOpen
  \bibfield  {author} {\bibinfo {author} {\bibfnamefont {Zhao-Feng}\
  \bibnamefont {Ding}}, \bibinfo {author} {\bibfnamefont {Yan-Xing}\
  \bibnamefont {Yang}}, \bibinfo {author} {\bibfnamefont {Jian}\ \bibnamefont
  {Zhang}}, \bibinfo {author} {\bibfnamefont {Cheng}\ \bibnamefont {Tan}},
  \bibinfo {author} {\bibfnamefont {Zi-Hao}\ \bibnamefont {Zhu}}, \bibinfo
  {author} {\bibfnamefont {Gang}\ \bibnamefont {Chen}}, \ and\ \bibinfo
  {author} {\bibfnamefont {Lei}\ \bibnamefont {Shu}},\ }\bibfield  {title}
  {\enquote {\bibinfo {title} {{Possible gapless spin liquid in the rare-earth
  kagome lattice magnet Tm$_3$Sb$_3$Zn$_2$O$_{14}$}},}\ }\href {\doibase
  10.1103/PhysRevB.98.174404} {\bibfield  {journal} {\bibinfo  {journal} {Phys.
  Rev. B}\ }\textbf {\bibinfo {volume} {98}},\ \bibinfo {pages} {174404}
  (\bibinfo {year} {2018})}\BibitemShut {NoStop}%
\bibitem [{\citenamefont {Toby}(2001)}]{Toby:hw0089}%
  \BibitemOpen
  \bibfield  {author} {\bibinfo {author} {\bibfnamefont {Brian~H.}\
  \bibnamefont {Toby}},\ }\bibfield  {title} {\enquote {\bibinfo {title}
  {Expgui, a graphical user interface for gsas},}\ }\href {\doibase
  10.1107/S0021889801002242} {\bibfield  {journal} {\bibinfo  {journal}
  {J. Appl. Crystallogr.}\ }\textbf {\bibinfo {volume} {34}},\
  \bibinfo {pages} {210} (\bibinfo {year} {2001})}\BibitemShut {NoStop}%
\bibitem [{\citenamefont {Toby}\ and\ \citenamefont
  {Von~Dreele}(2013)}]{Toby:aj5212}%
  \BibitemOpen
  \bibfield  {author} {\bibinfo {author} {\bibfnamefont {Brian~H.}\
  \bibnamefont {Toby}}\ and\ \bibinfo {author} {\bibfnamefont {Robert~B.}\
  \bibnamefont {Von~Dreele}},\ }\bibfield  {title} {\enquote {\bibinfo {title}
  {Gsas-ii: the genesis of a modern open-source all purpose crystallography
  software package},}\ }\href {\doibase 10.1107/S0021889813003531} {\bibfield
  {journal} {\bibinfo  {journal} {J. Appl. Crystallogr.}\ }\textbf
  {\bibinfo {volume} {46}},\ \bibinfo {pages} {544} (\bibinfo {year}
  {2013})}\BibitemShut {NoStop}%
\bibitem [{\citenamefont {Bert}\ \emph {et~al.}(2007)\citenamefont {Bert},
  \citenamefont {Nakamae}, \citenamefont {Ladieu}, \citenamefont {L'H\^ote},
  \citenamefont {Bonville}, \citenamefont {Duc}, \citenamefont {Trombe},\ and\
  \citenamefont {Mendels}}]{PhysRevB.76.132411}%
  \BibitemOpen
  \bibfield  {author} {\bibinfo {author} {\bibfnamefont {F.}~\bibnamefont
  {Bert}}, \bibinfo {author} {\bibfnamefont {S.}~\bibnamefont {Nakamae}},
  \bibinfo {author} {\bibfnamefont {F.}~\bibnamefont {Ladieu}}, \bibinfo
  {author} {\bibfnamefont {D.}~\bibnamefont {L'H\^ote}}, \bibinfo {author}
  {\bibfnamefont {P.}~\bibnamefont {Bonville}}, \bibinfo {author}
  {\bibfnamefont {F.}~\bibnamefont {Duc}}, \bibinfo {author} {\bibfnamefont
  {J.-C.}\ \bibnamefont {Trombe}}, \ and\ \bibinfo {author} {\bibfnamefont
  {P.}~\bibnamefont {Mendels}},\ }\bibfield  {title} {\enquote {\bibinfo
  {title} {{Low temperature magnetization of the $S=\frac{1}{2}$ kagome
  antiferromagnet ZnCu$_3$(OH)$_6$Cl$_2$}},}\ }\href {\doibase
  10.1103/PhysRevB.76.132411} {\bibfield  {journal} {\bibinfo  {journal} {Phys.
  Rev. B}\ }\textbf {\bibinfo {volume} {76}},\ \bibinfo {pages} {132411}
  (\bibinfo {year} {2007})}\BibitemShut {NoStop}%
\bibitem [{\citenamefont {Balz}\ \emph {et~al.}(2017)\citenamefont {Balz},
  \citenamefont {Lake}, \citenamefont {Nazmul~Islam}, \citenamefont {Singh},
  \citenamefont {Rodriguez-Rivera}, \citenamefont {Guidi}, \citenamefont
  {Wheeler}, \citenamefont {Simeoni},\ and\ \citenamefont
  {Ryll}}]{PhysRevB.95.174414}%
  \BibitemOpen
  \bibfield  {author} {\bibinfo {author} {\bibfnamefont {Christian}\
  \bibnamefont {Balz}}, \bibinfo {author} {\bibfnamefont {Bella}\ \bibnamefont
  {Lake}}, \bibinfo {author} {\bibfnamefont {A.~T.~M.}\ \bibnamefont
  {Nazmul~Islam}}, \bibinfo {author} {\bibfnamefont {Yogesh}\ \bibnamefont
  {Singh}}, \bibinfo {author} {\bibfnamefont {Jose~A.}\ \bibnamefont
  {Rodriguez-Rivera}}, \bibinfo {author} {\bibfnamefont {Tatiana}\ \bibnamefont
  {Guidi}}, \bibinfo {author} {\bibfnamefont {Elisa~M.}\ \bibnamefont
  {Wheeler}}, \bibinfo {author} {\bibfnamefont {Giovanna~G.}\ \bibnamefont
  {Simeoni}}, \ and\ \bibinfo {author} {\bibfnamefont {Hanjo}\ \bibnamefont
  {Ryll}},\ }\bibfield  {title} {\enquote {\bibinfo {title} {{Magnetic
  Hamiltonian and phase diagram of the quantum spin liquid
  Ca$_{10}$Cr$_7$O$_{28}$}},}\ }\href {\doibase 10.1103/PhysRevB.95.174414}
  {\bibfield  {journal} {\bibinfo  {journal} {Phys. Rev. B}\ }\textbf {\bibinfo
  {volume} {95}},\ \bibinfo {pages} {174414} (\bibinfo {year}
  {2017})}\BibitemShut {NoStop}%
\bibitem [{\citenamefont {Scheie}\ \emph {et~al.}(2018)\citenamefont {Scheie},
  \citenamefont {Sanders}, \citenamefont {Krizan}, \citenamefont
  {Christianson}, \citenamefont {Garlea}, \citenamefont {Cava},\ and\
  \citenamefont {Broholm}}]{PhysRevB.98.134401}%
  \BibitemOpen
  \bibfield  {author} {\bibinfo {author} {\bibfnamefont {A.}~\bibnamefont
  {Scheie}}, \bibinfo {author} {\bibfnamefont {M.}~\bibnamefont {Sanders}},
  \bibinfo {author} {\bibfnamefont {J.}~\bibnamefont {Krizan}}, \bibinfo
  {author} {\bibfnamefont {A.~D.}\ \bibnamefont {Christianson}}, \bibinfo
  {author} {\bibfnamefont {V.~O.}\ \bibnamefont {Garlea}}, \bibinfo {author}
  {\bibfnamefont {R.~J.}\ \bibnamefont {Cava}}, \ and\ \bibinfo {author}
  {\bibfnamefont {C.}~\bibnamefont {Broholm}},\ }\bibfield  {title} {\enquote
  {\bibinfo {title} {{Crystal field levels and magnetic anisotropy in the
  kagome compounds Nd$_3$Sb$_3$Mg$_2$O$_{14}$, Nd$_3$Sb$_3$Zn$_2$O$_{14}$, and
  Pr$_3$Sb$_3$Mg$_2$O$_{14}$}},}\ }\href {\doibase 10.1103/PhysRevB.98.134401}
  {\bibfield  {journal} {\bibinfo  {journal} {Phys. Rev. B}\ }\textbf {\bibinfo
  {volume} {98}},\ \bibinfo {pages} {134401} (\bibinfo {year}
  {2018})}\BibitemShut {NoStop}%
\bibitem [{\citenamefont {Hutchings}(1964)}]{HUTCHINGS1964227}%
  \BibitemOpen
  \bibfield  {author} {\bibinfo {author} {\bibfnamefont {M.T.}\ \bibnamefont
  {Hutchings}},\ }\bibfield  {title} {\enquote {\bibinfo {title} {Point-charge
  calculations of energy levels of magnetic ions in crystalline electric
  fields},}\ \
  }(\bibinfo  {publisher} {Academic Press},\ \bibinfo {year} {1964})\ pp.\
  \bibinfo {pages} {227 -- 273}\BibitemShut {NoStop}%
\bibitem [{\citenamefont {Stevens}(1952)}]{Stevens_1952}%
  \BibitemOpen
  \bibfield  {author} {\bibinfo {author} {\bibfnamefont {K~W~H}\ \bibnamefont
  {Stevens}},\ }\bibfield  {title} {\enquote {\bibinfo {title} {Matrix elements
  and operator equivalents connected with the magnetic properties of rare earth
  ions},}\ }\href {\doibase 10.1088/0370-1298/65/3/308} {\bibfield  {journal}
  {\bibinfo  {journal} {P. Phys. Soc. A}\
  }\textbf {\bibinfo {volume} {65}},\ \bibinfo {pages} {209} (\bibinfo
  {year} {1952})}\BibitemShut {NoStop}%
\bibitem [{\citenamefont {Dun}\ \emph {et~al.}(2020{\natexlab{a}})\citenamefont
  {Dun}, \citenamefont {Bai}, \citenamefont {Paddison}, \citenamefont
  {Hollingworth}, \citenamefont {Butch}, \citenamefont {Cruz}, \citenamefont
  {Stone}, \citenamefont {Hong}, \citenamefont {Demmel}, \citenamefont
  {Mourigal},\ and\ \citenamefont {Zhou}}]{PhysRevX.10.031069}%
  \BibitemOpen
  \bibfield  {author} {\bibinfo {author} {\bibfnamefont {Zhiling}\ \bibnamefont
  {Dun}}, \bibinfo {author} {\bibfnamefont {Xiaojian}\ \bibnamefont {Bai}},
  \bibinfo {author} {\bibfnamefont {Joseph A.~M.}\ \bibnamefont {Paddison}},
  \bibinfo {author} {\bibfnamefont {Emily}\ \bibnamefont {Hollingworth}},
  \bibinfo {author} {\bibfnamefont {Nicholas~P.}\ \bibnamefont {Butch}},
  \bibinfo {author} {\bibfnamefont {Clarina~D.}\ \bibnamefont {Cruz}}, \bibinfo
  {author} {\bibfnamefont {Matthew~B.}\ \bibnamefont {Stone}}, \bibinfo
  {author} {\bibfnamefont {Tao}\ \bibnamefont {Hong}}, \bibinfo {author}
  {\bibfnamefont {Franz}\ \bibnamefont {Demmel}}, \bibinfo {author}
  {\bibfnamefont {Martin}\ \bibnamefont {Mourigal}}, \ and\ \bibinfo {author}
  {\bibfnamefont {Haidong}\ \bibnamefont {Zhou}},\ }\bibfield  {title}
  {\enquote {\bibinfo {title} {{Quantum Versus Classical Spin Fragmentation in
  Dipolar Kagome Ice
  ${\mathrm{Ho}}_{3}{\mathrm{Mg}}_{2}{\mathrm{Sb}}_{3}{\mathrm{O}}_{14}$}},}\
  }\href {\doibase 10.1103/PhysRevX.10.031069} {\bibfield  {journal} {\bibinfo
  {journal} {Phys. Rev. X}\ }\textbf {\bibinfo {volume} {10}},\ \bibinfo
  {pages} {031069} (\bibinfo {year} {2020}{\natexlab{a}})}\BibitemShut
  {NoStop}%
\bibitem [{\citenamefont {Dun}\ \emph {et~al.}(2020{\natexlab{a}})\citenamefont
  {Dun}, \citenamefont {Bai}, \citenamefont {Stone}, \citenamefont {Zhou},\
  and\ \citenamefont {Mourigal}}]{dun2020effective}%
  \BibitemOpen
  \bibfield  {author} {\bibinfo {author} {\bibfnamefont {Zhiling}\ \bibnamefont
  {Dun}}, \bibinfo {author} {\bibfnamefont {Xiaojian}\ \bibnamefont {Bai}},
  \bibinfo {author} {\bibfnamefont {Matthew~B.}\ \bibnamefont {Stone}},
  \bibinfo {author} {\bibfnamefont {Haidong}\ \bibnamefont {Zhou}}, \ and\
  \bibinfo {author} {\bibfnamefont {Martin}\ \bibnamefont {Mourigal}},\
  }\href@noop {} {\enquote {\bibinfo {title} {Effective point-charge analysis
  of crystal electric fields -- application to rare-earth pyrochlores and
  tripod kagome magnets R$_3$Mg$_2$Sb$_3$O$_{14}$},}\ } \ \Eprint {http://arxiv.org/abs/2004.10957}
  {arXiv:2004.10957} \BibitemShut {NoStop}%
\bibitem [{\citenamefont {Li}\ \emph {et~al.}(2020)\citenamefont {Li},
  \citenamefont {Bachus}, \citenamefont {Deng}, \citenamefont {Schmidt},
  \citenamefont {Thoma}, \citenamefont {Hutanu}, \citenamefont {Tokiwa},
  \citenamefont {Tsirlin},\ and\ \citenamefont
  {Gegenwart}}]{PhysRevX.10.011007}%
  \BibitemOpen
  \bibfield  {author} {\bibinfo {author} {\bibfnamefont {Yuesheng}\
  \bibnamefont {Li}}, \bibinfo {author} {\bibfnamefont {Sebastian}\
  \bibnamefont {Bachus}}, \bibinfo {author} {\bibfnamefont {Hao}\ \bibnamefont
  {Deng}}, \bibinfo {author} {\bibfnamefont {Wolfgang}\ \bibnamefont
  {Schmidt}}, \bibinfo {author} {\bibfnamefont {Henrik}\ \bibnamefont {Thoma}},
  \bibinfo {author} {\bibfnamefont {Vladimir}\ \bibnamefont {Hutanu}}, \bibinfo
  {author} {\bibfnamefont {Yoshifumi}\ \bibnamefont {Tokiwa}}, \bibinfo
  {author} {\bibfnamefont {Alexander~A.}\ \bibnamefont {Tsirlin}}, \ and\
  \bibinfo {author} {\bibfnamefont {Philipp}\ \bibnamefont {Gegenwart}},\
  }\bibfield  {title} {\enquote {\bibinfo {title} {{Partial Up-Up-Down Order
  with the Continuously Distributed Order Parameter in the Triangular
  Antiferromagnet TmMgGaO$_4$}},}\ }\href {\doibase 10.1103/PhysRevX.10.011007}
  {\bibfield  {journal} {\bibinfo  {journal} {Phys. Rev. X}\ }\textbf {\bibinfo
  {volume} {10}},\ \bibinfo {pages} {011007} (\bibinfo {year}
  {2020})}\BibitemShut {NoStop}%
\bibitem [{\citenamefont {Yu}\ \emph {et~al.}(2013)\citenamefont {Yu},
  \citenamefont {Mole}, \citenamefont {Noakes}, \citenamefont {Kennedy},\ and\
  \citenamefont {Robinson}}]{doi:10.7566/JPSJS.82SA.SA027}%
  \BibitemOpen
  \bibfield  {author} {\bibinfo {author} {\bibfnamefont {Dehong}\ \bibnamefont
  {Yu}}, \bibinfo {author} {\bibfnamefont {Richard}\ \bibnamefont {Mole}},
  \bibinfo {author} {\bibfnamefont {Terry}\ \bibnamefont {Noakes}}, \bibinfo
  {author} {\bibfnamefont {Shane}\ \bibnamefont {Kennedy}}, \ and\ \bibinfo
  {author} {\bibfnamefont {Robert}\ \bibnamefont {Robinson}},\ }\bibfield
  {title} {\enquote {\bibinfo {title} {{Pelican---a Time of Flight Cold Neutron
  Polarization Analysis Spectrometer at OPAL}},}\ }\href {\doibase
  10.7566/JPSJS.82SA.SA027} {\bibfield  {journal} {\bibinfo  {journal} {J. Phys. Soc. Jpn.}\ }\textbf {\bibinfo {volume} {82}},\
  \bibinfo {pages} {SA027} (\bibinfo {year} {2013})}\BibitemShut {NoStop}%
\bibitem [{\citenamefont {Clark}\ \emph {et~al.}(2019)\citenamefont {Clark},
  \citenamefont {Sala}, \citenamefont {Maharaj}, \citenamefont {Stone},
  \citenamefont {Knight}, \citenamefont {Telling}, \citenamefont {Wang},
  \citenamefont {Xu}, \citenamefont {Kim}, \citenamefont {Li}, \citenamefont
  {Cheong},\ and\ \citenamefont {Gaulin}}]{np15.262}%
  \BibitemOpen
  \bibfield  {author} {\bibinfo {author} {\bibfnamefont {Lucy}\ \bibnamefont
  {Clark}}, \bibinfo {author} {\bibfnamefont {Gabriele}\ \bibnamefont {Sala}},
  \bibinfo {author} {\bibfnamefont {Dalini~D.}\ \bibnamefont {Maharaj}},
  \bibinfo {author} {\bibfnamefont {Matthew~B.}\ \bibnamefont {Stone}},
  \bibinfo {author} {\bibfnamefont {Kevin~S.}\ \bibnamefont {Knight}}, \bibinfo
  {author} {\bibfnamefont {Mark T.~F.}\ \bibnamefont {Telling}}, \bibinfo
  {author} {\bibfnamefont {Xueyun}\ \bibnamefont {Wang}}, \bibinfo {author}
  {\bibfnamefont {Xianghan}\ \bibnamefont {Xu}}, \bibinfo {author}
  {\bibfnamefont {Jaewook}\ \bibnamefont {Kim}}, \bibinfo {author}
  {\bibfnamefont {Yanbin}\ \bibnamefont {Li}}, \bibinfo {author} {\bibfnamefont
  {Sang-Wook}\ \bibnamefont {Cheong}}, \ and\ \bibinfo {author} {\bibfnamefont
  {Bruce~D.}\ \bibnamefont {Gaulin}},\ }\bibfield  {title} {\enquote {\bibinfo
  {title} {{Two-dimensional spin liquid behaviour in the triangular-honeycomb
  antiferromagnet TbInO$_3$}},}\ }\href {\doibase
  https://doi.org/10.1038/s41567-018-0407-2} {\bibfield  {journal} {\bibinfo
  {journal} {Nat. Phys.}\ }\textbf {\bibinfo {volume} {15}},\ \bibinfo
  {pages} {262} (\bibinfo {year} {2019})}\BibitemShut {NoStop}%
\bibitem [{\citenamefont {Balz}\ \emph {et~al.}(2016)\citenamefont {Balz},
  \citenamefont {Lake}, \citenamefont {Reuther}, \citenamefont {Luetkens},
  \citenamefont {Schonemann}, \citenamefont {Herrmannsdorfer}, \citenamefont
  {Singh}, \citenamefont {Nazmul~Islam}, \citenamefont {Wheeler}, \citenamefont
  {Rodriguez-Rivera}, \citenamefont {Guidi}, \citenamefont {Simeoni},
  \citenamefont {Baines},\ and\ \citenamefont {Ryll}}]{np12_942}%
  \BibitemOpen
  \bibfield  {author} {\bibinfo {author} {\bibfnamefont {Christian}\
  \bibnamefont {Balz}}, \bibinfo {author} {\bibfnamefont {Bella}\ \bibnamefont
  {Lake}}, \bibinfo {author} {\bibfnamefont {Johannes}\ \bibnamefont
  {Reuther}}, \bibinfo {author} {\bibfnamefont {Hubertus}\ \bibnamefont
  {Luetkens}}, \bibinfo {author} {\bibfnamefont {Rico}\ \bibnamefont
  {Schonemann}}, \bibinfo {author} {\bibfnamefont {Thomas}\ \bibnamefont
  {Herrmannsdorfer}}, \bibinfo {author} {\bibfnamefont {Yogesh}\ \bibnamefont
  {Singh}}, \bibinfo {author} {\bibfnamefont {A.~T.~M.}\ \bibnamefont
  {Nazmul~Islam}}, \bibinfo {author} {\bibfnamefont {Elisa~M.}\ \bibnamefont
  {Wheeler}}, \bibinfo {author} {\bibfnamefont {Jose~A.}\ \bibnamefont
  {Rodriguez-Rivera}}, \bibinfo {author} {\bibfnamefont {Tatiana}\ \bibnamefont
  {Guidi}}, \bibinfo {author} {\bibfnamefont {Giovanna~G.}\ \bibnamefont
  {Simeoni}}, \bibinfo {author} {\bibfnamefont {Chris}\ \bibnamefont {Baines}},
  \ and\ \bibinfo {author} {\bibfnamefont {Hanjo}\ \bibnamefont {Ryll}},\
  }\bibfield  {title} {\enquote {\bibinfo {title} {Physical realization of a
  quantum spin liquid based on a complex frustration mechanism},}\ }\href@noop
  {} {\bibfield  {journal} {\bibinfo  {journal} {Nat. Phys.}\ }\textbf
  {\bibinfo {volume} {12}},\ \bibinfo {pages} {942} (\bibinfo {year}
  {2016})}\BibitemShut {NoStop}%
\bibitem [{\citenamefont {Gao}\ \emph {et~al.}(2019)\citenamefont {Gao},
  \citenamefont {Chen}, \citenamefont {Tam}, \citenamefont {Huang},
  \citenamefont {Sasmal}, \citenamefont {Adroja}, \citenamefont {Ye},
  \citenamefont {Cao}, \citenamefont {Sala}, \citenamefont {Stone},
  \citenamefont {Baines}, \citenamefont {Verezhak}, \citenamefont {Hu},
  \citenamefont {Chung}, \citenamefont {Xu}, \citenamefont {Cheong},
  \citenamefont {Nallaiyan}, \citenamefont {Spagna}, \citenamefont {Maple},
  \citenamefont {Nevidomskyy}, \citenamefont {Morosan}, \citenamefont {Chen},\
  and\ \citenamefont {Dai}}]{np15_1052}%
  \BibitemOpen
  \bibfield  {author} {\bibinfo {author} {\bibfnamefont {Bin}\ \bibnamefont
  {Gao}}, \bibinfo {author} {\bibfnamefont {Tong}\ \bibnamefont {Chen}},
  \bibinfo {author} {\bibfnamefont {David~W.}\ \bibnamefont {Tam}}, \bibinfo
  {author} {\bibfnamefont {Chien-Lung}\ \bibnamefont {Huang}}, \bibinfo
  {author} {\bibfnamefont {Kalyan}\ \bibnamefont {Sasmal}}, \bibinfo {author}
  {\bibfnamefont {Devashibhai~T.}\ \bibnamefont {Adroja}}, \bibinfo {author}
  {\bibfnamefont {Feng}\ \bibnamefont {Ye}}, \bibinfo {author} {\bibfnamefont
  {Huibo}\ \bibnamefont {Cao}}, \bibinfo {author} {\bibfnamefont {Gabriele}\
  \bibnamefont {Sala}}, \bibinfo {author} {\bibfnamefont {Matthew~B.}\
  \bibnamefont {Stone}}, \bibinfo {author} {\bibfnamefont {Christopher}\
  \bibnamefont {Baines}}, \bibinfo {author} {\bibfnamefont {Joel A.~T.}\
  \bibnamefont {Verezhak}}, \bibinfo {author} {\bibfnamefont {Haoyu}\
  \bibnamefont {Hu}}, \bibinfo {author} {\bibfnamefont {Jae-Ho}\ \bibnamefont
  {Chung}}, \bibinfo {author} {\bibfnamefont {Xianghan}\ \bibnamefont {Xu}},
  \bibinfo {author} {\bibfnamefont {Sang-Wook}\ \bibnamefont {Cheong}},
  \bibinfo {author} {\bibfnamefont {Manivannan}\ \bibnamefont {Nallaiyan}},
  \bibinfo {author} {\bibfnamefont {Stefano}\ \bibnamefont {Spagna}}, \bibinfo
  {author} {\bibfnamefont {M.~Brian}\ \bibnamefont {Maple}}, \bibinfo {author}
  {\bibfnamefont {Andriy~H.}\ \bibnamefont {Nevidomskyy}}, \bibinfo {author}
  {\bibfnamefont {Emilia}\ \bibnamefont {Morosan}}, \bibinfo {author}
  {\bibfnamefont {Gang}\ \bibnamefont {Chen}}, \ and\ \bibinfo {author}
  {\bibfnamefont {Pengcheng}\ \bibnamefont {Dai}},\ }\bibfield  {title}
  {\enquote {\bibinfo {title} {{Experimental signatures of a three-dimensional
  quantum spin liquid in effective spin-1/2 Ce$_2$Zr$_2$O$_7$ pyrochlore}},}\
  }\href@noop {} {\bibfield  {journal} {\bibinfo  {journal} {Nat. Phys.}\
  }\textbf {\bibinfo {volume} {15}},\ \bibinfo {pages} {1052} (\bibinfo
  {year} {2019})}\BibitemShut {NoStop}%
\bibitem [{\citenamefont {Furrer}\ and\ \citenamefont
  {Waldmann}(2013)}]{RevModPhys.85.367}%
  \BibitemOpen
  \bibfield  {author} {\bibinfo {author} {\bibfnamefont {Albert}\ \bibnamefont
  {Furrer}}\ and\ \bibinfo {author} {\bibfnamefont {Oliver}\ \bibnamefont
  {Waldmann}},\ }\bibfield  {title} {\enquote {\bibinfo {title} {Magnetic
  cluster excitations},}\ }\href {\doibase 10.1103/RevModPhys.85.367}
  {\bibfield  {journal} {\bibinfo  {journal} {Rev. Mod. Phys.}\ }\textbf
  {\bibinfo {volume} {85}},\ \bibinfo {pages} {367} (\bibinfo {year}
  {2013})}\BibitemShut {NoStop}%
\bibitem [{\citenamefont {Riedl}\ \emph {et~al.}(2019)\citenamefont {Riedl},
  \citenamefont {Valent\'{\i}},\ and\ \citenamefont {Winter}}]{nc10_2561}%
  \BibitemOpen
  \bibfield  {author} {\bibinfo {author} {\bibfnamefont {Kira}\ \bibnamefont
  {Riedl}}, \bibinfo {author} {\bibfnamefont {Roser}\ \bibnamefont
  {Valent\'{\i}}}, \ and\ \bibinfo {author} {\bibfnamefont {Stephen~M.}\
  \bibnamefont {Winter}},\ }\bibfield  {title} {\enquote {\bibinfo {title}
  {Critical spin liquid versus valence-bond glass in a triangular-lattice
  organic antiferromagnet},}\ }\href@noop {} {\bibfield  {journal} {\bibinfo
  {journal} {Nat. Commun.}\ }\textbf {\bibinfo {volume} {10}},\
  \bibinfo {pages} {2561} (\bibinfo {year} {2019})}\BibitemShut {NoStop}%
\bibitem [{\citenamefont {Pustogow}\ \emph {et~al.}(2020)\citenamefont
  {Pustogow}, \citenamefont {Le}, \citenamefont {Wang}, \citenamefont {Luo},
  \citenamefont {Gati}, \citenamefont {Schubert}, \citenamefont {Lang},\ and\
  \citenamefont {Brown}}]{PhysRevB.101.140401}%
  \BibitemOpen
  \bibfield  {author} {\bibinfo {author} {\bibfnamefont {A.}~\bibnamefont
  {Pustogow}}, \bibinfo {author} {\bibfnamefont {T.}~\bibnamefont {Le}},
  \bibinfo {author} {\bibfnamefont {H.-H.}\ \bibnamefont {Wang}}, \bibinfo
  {author} {\bibfnamefont {Yongkang}\ \bibnamefont {Luo}}, \bibinfo {author}
  {\bibfnamefont {E.}~\bibnamefont {Gati}}, \bibinfo {author} {\bibfnamefont
  {H.}~\bibnamefont {Schubert}}, \bibinfo {author} {\bibfnamefont
  {M.}~\bibnamefont {Lang}}, \ and\ \bibinfo {author} {\bibfnamefont {S.~E.}\
  \bibnamefont {Brown}},\ }\bibfield  {title} {\enquote {\bibinfo {title}
  {Impurity moments conceal low-energy relaxation of quantum spin liquids},}\
  }\href {\doibase 10.1103/PhysRevB.101.140401} {\bibfield  {journal} {\bibinfo
   {journal} {Phys. Rev. B}\ }\textbf {\bibinfo {volume} {101}},\ \bibinfo
  {pages} {140401} (\bibinfo {year} {2020})}\BibitemShut {NoStop}%
%\bibitem [{\citenamefont {Dun}\ \emph {et~al.}(2020{\natexlab{b}})\citenamefont
 % {Dun}, \citenamefont {Bai}, \citenamefont {Stone}, \citenamefont {Zhou},\
 % and\ \citenamefont {Mourigal}}]{arXiv:2004.10957}%
 % \BibitemOpen
  %\bibfield  {author} {\bibinfo {author} {\bibfnamefont {Zhiling}\ \bibnamefont
  %{Dun}}, \bibinfo {author} {\bibfnamefont {Xiaojian}\ \bibnamefont {Bai}},
  %\bibinfo {author} {\bibfnamefont {Matthew~B.}\ \bibnamefont {Stone}},
 % \bibinfo {author} {\bibfnamefont {Haidong}\ \bibnamefont {Zhou}}, \ and\
 % \bibinfo {author} {\bibfnamefont {Martin}\ \bibnamefont {Mourigal}},\
 % }\bibfield  {title} {\enquote {\bibinfo {title} {{Effective point-charge
 % analysis of crystal electric fields -- application to rare-earth pyrochlores
 % and tripod kagome magnets $R_3$Mg$_2$Sb$_3$O$_{14}$}},}\ }\href@noop {}
 % {\bibfield  {journal} {\bibinfo  {journal} {arXiv:2004.10957}\ }}\BibitemShut {NoStop}%
\end{thebibliography}


\end{document}